\documentclass[twocolumn,prc,aps,showpacs,amsmath,amssymb]{revtex4}
\usepackage{graphicx}
 \newcommand\la{\langle}
 \newcommand\ra{\rangle}
 \newcommand\beq{\begin{equation}}
 
 \newcommand\eeq{\end{equation}}
 \newcommand\beqn{\begin{eqnarray}}
 \newcommand\eeqn{\end{eqnarray}}
 \newcommand\GeV{{\rm GeV}}
 \newcommand{\doublespace} {
 \renewcommand{\baselinestretch} {1.6}
\large\normalsize}
\def\doublespace{\def\baselinestretch{1.6}\large\normalsize}
\def\normalspace{\def\baselinestretch{1.0}\normalsize}

\def\Caption#1{
  \normalspace
  \begin{quotation}\caption{\sl #1}\end{quotation}
  \doublespace
}
\def\im{\mbox{Im}\,}
\def\re{\mbox{Re}\,}
\def\mb{\,\mbox{mb}}
\def\fm{\,\mbox{fm}}
\def\GeV{\,\mbox{GeV}}
\def\TeV{\,\mbox{TeV}}

\def\Pom{{\rm I\!P}}

\def\lsim{\mathrel{\rlap{\lower4pt\hbox{\hskip1pt$\sim$}}
    \raise1pt\hbox{$<$}}}         %less than or approx. symbol
\def\gsim{\mathrel{\rlap{\lower4pt\hbox{\hskip1pt$\sim$}}
    \raise1pt\hbox{$>$}}}         %greater than or approx. symbol
\def\la{\langle}
\def\ra{\rangle}

\begin{document}

\title{Pion-pion cross section from proton-proton collisions at the LHC}

\author{B. Z. Kopeliovich}
\author{ I. K. Potashnikova}
\author{Iv\'an Schmidt}

\affiliation{\centerline{$^1$Departamento de F\'{\i}sica,
Universidad T\'ecnica Federico Santa Mar\'{\i}a; and}
Centro Cient\'ifico-Tecnol\'ogico de Valpara\'iso;
Casilla 110-V, Valpara\'iso, Chile}

\author{H. J. Pirner$^{1}$}
\author{K. Reygers$^{2}$}

\affiliation{
{$^{1}$Institute for Theoretical Physics, University of Heidelberg, Germany
}\\
{$^2$Physikalisches Institut, University of Heidelberg, Germany}}

\begin{abstract}

The zero-degree calorimeters (ZDC)  installed in the ALICE, ATLAS and CMS experiments at the LHC, make possible simultaneous detection of forward-backward leading neutrons, $pp\to n\ X\ n$. Such data with sufficiently high statistics could be a source of information about the pion-pion total cross section at high energies, provided that the absorption corrections, which are expected to be strong, are well understood. Otherwise, making a plausible assumption about the magnitude of the pion-pion cross section, one can consider such measurements as a way to study the absorption effects, which is the main focus of the present paper. These effects introduced at the amplitude level, are found to be different for the pion fluxes, which either conserve or flip the nucleon helicity. The pion fluxes from both colliding protons are essentially reduced by absorption, moreover, there is a common absorption suppression factor, which breaks down the factorized form of the cross section. 
We also evaluate the feed-down corrections related to the initial/final state inelastic processes possessing a rapidity gap, and found them to be small in the kinematic range under consideration. The contribution of other iso-vector Reggeons,
spin-flip natural parity $\rho$ and $a_2$, and spin non-flip unnatural parity $a_1$ are also evaluated and found to be rather small. 
\end{abstract}

%\date{\today}


\maketitle

\section{Introduction}

The proton-proton elastic scattering cross section has been measured in  a wide range of  energies, and recently up to the highest energy of the LHC, $\sqrt{s}=7\TeV$  \cite{totem,alpha}. At the same time,
measurements of the pion-nucleon cross section have been restricted so far to rather low energies, up to about $\sqrt{s}=35\GeV$ \cite{pdg}. The pion-pion cross section cannot be measured directly, and has been extracted from data only at very low energies near  threshold \cite{review}.
The theoretical description of elastic scattering has been based so far only on phenomenological
models. Even the simplest versions of Regge models, assuming Pomeron pole dominance (no cuts) \cite{landshoff} still describe the available data reasonably well, in spite of the obvious
problems with the unitarity bound at higher energies. Among the unitarized models \cite{gotsman,soffer,erasmo,martin,kaidalov,ostapchenko}, a precise prediction of the elastic cross section at the LHC was done
in \cite{k3p,kpp2012}. In contrast to the models treating the Pomeron as a simple Regge pole, an increasing rate of the energy dependence was predicted. Even a steeper rise of the cross sections at high energies is expected for   $\pi$-$p$ and $\pi$-$\pi$ scattering.
The models \cite{pir-1} based on non perturbative interaction dynamics fixed at low energies, predict an increasing cross section with energy. These models provided predictions for $pp$, $\pi p$ and $\pi\pi$ cross sections.

The possibility of having a pion-pion collider does not seem to be realistic, and it has not been seriously considered so far. However, one can make use of virtual pion beams. Indeed, nucleons are known to have pion clouds, with low virtuality, so high energy proton beams are accompanied 
by an intensive flux of high-energy pions, which participate in collisions. This way to measure electron-pion collisions was employed in the ZEUS \cite{zeus} and H1 \cite{h1} experiments at HERA. Pion contribution was singled out by detecting leading neutrons with large fractional momentum, $z$.
The main objective of these measurements was the determination of the pion structure function 
$F_2^\pi(x,Q^2)$ at low $x$. This task turned out to be not straightforward, because of
absorptive corrections, which suppress the cross section. In fact, recent study of these effects 
\cite{kpps-dis} found them to be quite strong, reducing significantly the cross section. A good description of data was achieved. A weaker effect of absorption was expected in Refs.~\cite{ap,strong2,kkmr,kmr}.

Detecting leading neutrons with large $z$ in $pp$ collisions one can access the $\pi$-$p$ total cross section
at energies much higher than with real pion beams. Apparently, the absorptive corrections in this case should be similar or stronger than in $\gamma^*$-$p$ collisions.
A detailed study of these effects was performed in \cite{kpss}.  However, no data from modern colliders have been available so far, except for a few points with large error bars from the PHENIX experiment \cite{phenix-neutrons1,phenix-neutrons2} and old data from ISR \cite{isr}. The normalization of the latter was found unreliable in \cite{kpss}.

Earlier attempts to extract the $\pi\pi$ and $\pi p$ cross sections from neutron production at low energies of fixed target experiments were made in \cite{hanlon1,hanlon2}, although no absorptive corrections were introduced. An attempt to calculate the effect of absorption for there processes were made in \cite{petrov,petrov2}, however only the shadow of the interaction between the colliding protons was included, while the main source of absorption was missed. The latter comes from the higher Fock components of the projectile proton, which contain a color octet-octet dipole as was demobsreated in \cite{kpss,kpps-dis}.
This explains in particular the observed cross section of leading neutron production in deep-inelastic scattering, which is independent of $Q^2$ \cite{zeus,h1,kpps-dis}. On the contrary, if the absorption effects were caused by $\gamma^*-p$ interactions in accordance with \cite{petrov},  then the fractional cross section of neutron production would steeply vary with $Q^2$ (see in \cite{kpps-dis}). 

The experiments ATLAS, CMS and ALICE at the LHC, are equipped with zero-degree calorimeters, which are able to detect neutrons at very small angles. This is ideal
for experimenting with pions accompanying the colliding protons. In particular, detecting leading neutrons, simultaneously produced in both directions, one can accesses pion-pion collisions 
at high c.m. energy, $s_{\pi\pi}=(1-z_1)(1-z_2)s$, where $z_{1,2}$ are the fractional momenta of the detected neutrons. Naturally, this process is also subject to strong absorptive corrections, which have not been studied so far. Our objective in this paper is to calculate these corrections, which would allow to extract pion-pion total cross section from the process $pp\to
nXn$ with two forward/backward neutrons detected with large $z_{1,2}$.

The paper is organized as follows. In Sect.~\ref{born} we describe the kinematics for the double-leading neutron production,
and the cross section in the Born approximation, i.e. without any absorptive corrections.
Sect.~\ref{absorption} is devoted to calculations of the absorptive corrections. This is done by switching the amplitude to impact parameter representation, where the absorptive corrections factorise (Sect.~\ref{impact}), and then coming back to momentum representation. The absorption suppression factors, also called gap survival amplitudes, are evaluated in Sect.~{\ref{survival}.
They originate from different types of initial/final state absorption effects. Interaction with the produced multi-particle system $X$ is described in Sect.~\ref{S1,2} and \ref{feed-down}, while
the effects of interaction between the spectator nucleons is  considered in Sect.~\ref{S(NN)}.
Factorization of the cross section into the product of two pion fluxes turns out to be heavily broken by $NN$-absorption effects.

In addition to pions, other iso-triplet Reggeons, $\rho,\ a_2$, and $a_1$ also contribute to neutron production. This background is evaluated in Sect.~\ref{reggeons}.

\section{Double-leading neutrons in pp collisions} \label{born}

The double rapidity gap process with two leading neutrons in the final state,
\beq
p+p\to n+X+n,
\label{100} 
\eeq
where both neutrons are produced with large fractional light-cone momenta $z_1$ and $z_2$,
can naturally be interpreted as a collision of two pion fluxes from the colliding protons, as is illustrated in fig.~\ref{fig:pi-pi}. 
%%%%%%%%%%%%%%%%%%%%%
\begin{figure}[htb]
\centerline{
  \scalebox{0.3}{\includegraphics{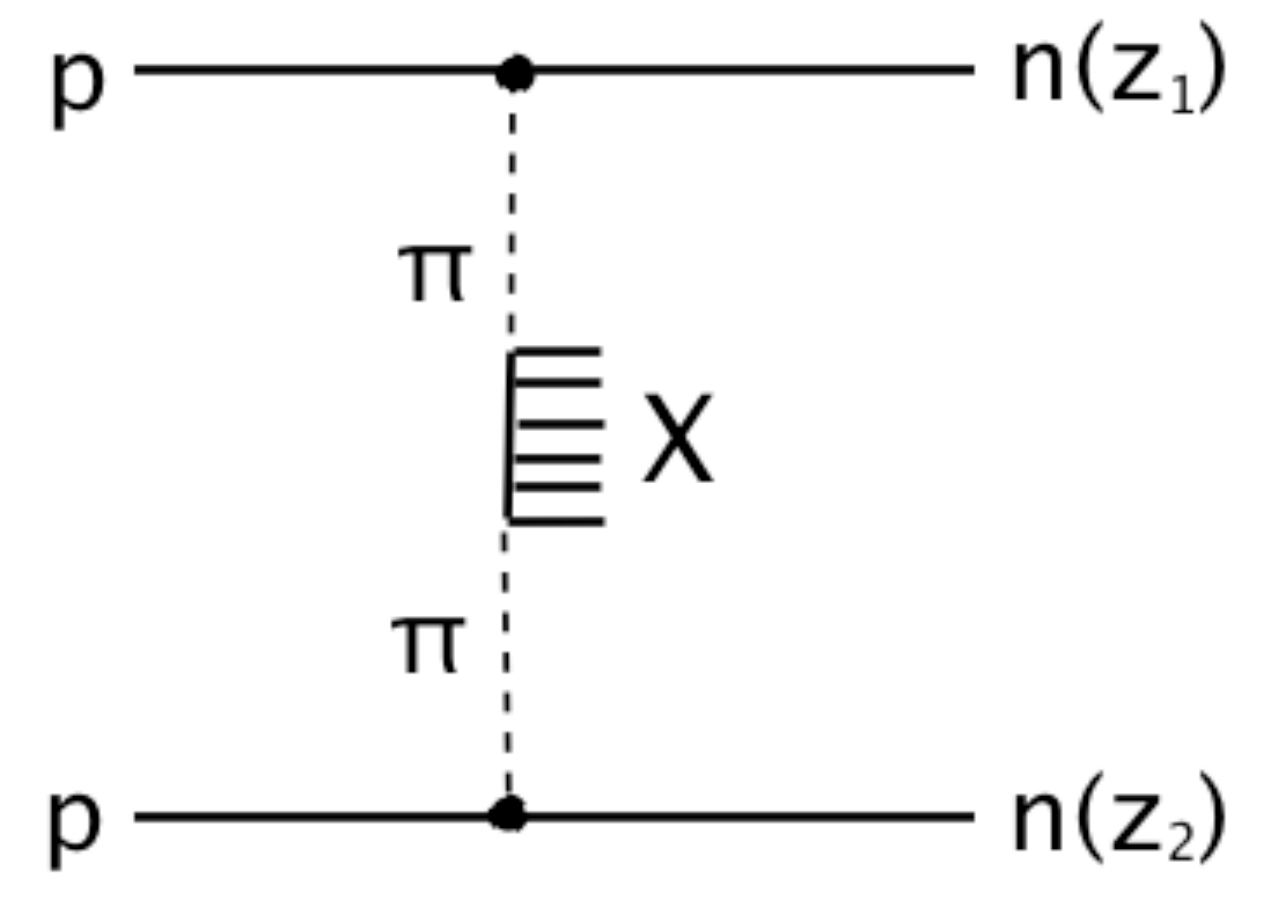}}}
\caption{\label{fig:pi-pi} Graphical representation for double neutron production with large $z$ in  $pp\to nXn$. }
 \end{figure}
%%%%%%%%%%%%%%%%%%%%%

The invariant mass of $X$, i.e. the $\pi\pi$ c.m. energy squared, is related to that of the $pp$ as,
\beq
\frac{s_{\pi\pi}}{s}\equiv\tau=(1-z_1)(1-z_2),
\label{120}
\eeq

Besides, the produced neutrons are characterised by  transverse momenta $\vec q_{i}$ and
4-momenta squared 
\beq
-t_i={1\over z_i}\left[\vec q_{i}^{\,2}+(1-z_i)^2m_N^2\right], 
\label{140}
\eeq
where $i=1,2$.

The cross section of the process (\ref{100}) in the Born approximation (no absorptive corrections) can be presented in the form,
 \beqn
\frac{d\sigma^B(pp\to nXn)}{dz_1dz_2\,dq_{1}^2dq_{2}^2}
&=& 
f^B_{\pi^+/p}(z_1,q_{1})\,
\sigma^{\pi^+\pi^+}_{tot}(\tau s)
\nonumber\\&\times& 
f^B_{\pi^+/p}(z_2,q_{2}),
\label{160}
\eeqn
where the pion flux in the proton (also called the pion distribution function) with fractional momentum $1-z$ reads \cite{kpp},
\beqn
f^B_{\pi^+/p}(z,q)&=&
-t\,G_{\pi^+pn}^2(t)
\left|\frac{\alpha_\pi^\prime\eta_\pi(t))}{8}\right|^2
\nonumber\\ &\times&
{1\over z}\,(1-z)^{1-2\alpha_\pi(t)}.
\label{180}
 \eeqn
 Here $\eta_\pi(t)$ is the phase factor, which can be expanded near
the pion pole as,
 \beq
\eta_\pi(t)=i-\cot\left[\frac{\pi\alpha_\pi(t)}{2}\right]\approx
i+\frac{2}{\pi\alpha_\pi^\prime}\,
\frac{1}{m_\pi^2-t}\,.
\label{190}
 \eeq
We neglect the small imaginary part in what follows.

The pion Regge trajectory is assumed to be linear,
$\alpha_\pi(t)=\alpha_\pi^\prime(t-m_\pi^2)$,
with $\alpha_\pi^\prime\approx 0.9\GeV^{-2}$.
The effective vertex function
$G_{\pi^+pn}(t)=g_{\pi^+pn}\exp(R_\pi^2t)$, where  
$g^2_{\pi^+pn}(t)/8\pi=13.85$. For further calculations we fix $R_\pi^2=0.3\GeV^{-2}$, which was adjusted to data and chosen in \cite{kpp,kaidalov1,kaidalov2,kkmr,kmr}
 as the most reliable value.

As an example, we calculate the flux $f_{\pi^+/p}^{B}(z,q)$ at $q=0$, plotted by a dashed curve in Fig.~\ref{fig:f0}. 
%%%%%%%%%%%%%%%%%%%%%
  \begin{figure}[htb]
\centerline{
  \scalebox{0.3}{\includegraphics{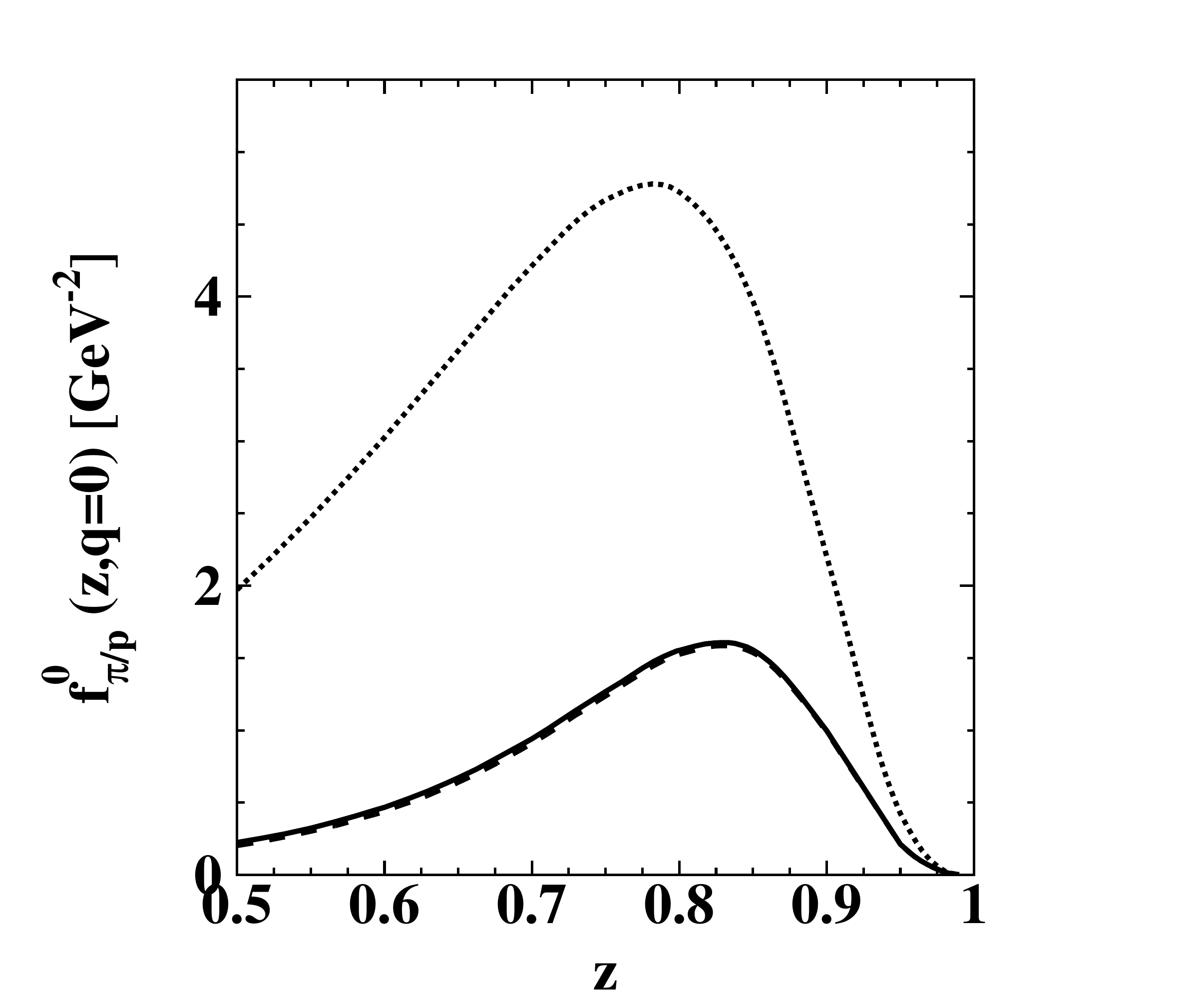}}}
 \caption{The forward flux of pions $f_{\pi^+/p}^{(0)}(z,q)$ at $q=0$,
 calculated in the Born approximation with Eq.~(\ref{274}) and including absorption, Eq.~(\ref{540}), plotted by dotted and dashed curves respectively, and by solid curve after adding the feed-down corrections.
 }
 \label{fig:f0}
 \end{figure}
 %%%%%%%%%%%%%%%%%%%%%

In experiments with a sufficiently large aperture one can accept all the leading neutrons
and rely on the cross section integrated over transverse momenta,
 \beq
\frac{d\sigma^B(pp\to nXn)}{dz_1dz_2}
=
F^B_{\pi^+\!/p}(z_1)
\sigma^{\pi^+\pi^+}_{tot}(\tau s)
F^B_{\pi^+\!/p}(z_2),
\label{195}
\eeq
where the $q$-integrated flux reads,
\beq
F^B_{\pi^+/p}(z)=-z\int\limits_{q_L^2}^\infty dt\,
 f^B_{\pi^+/p}(z,q);
 \label{197}
 \eeq 
 and
  \beq
q_L=\frac{1-z}{\sqrt{z}}\,m_N.
\label{220}
 \eeq

The  $q$-integrated pion flux $F^B_{\pi^+\!/p}(z)$ is plotted in Fig.~\ref{fig:F-B}
as function of neutron fractional momentum $z$.
%%%%%%%%%%%%%%%%%%%%%
\begin{figure}[htb]
\centerline{
  \scalebox{0.3}{\includegraphics{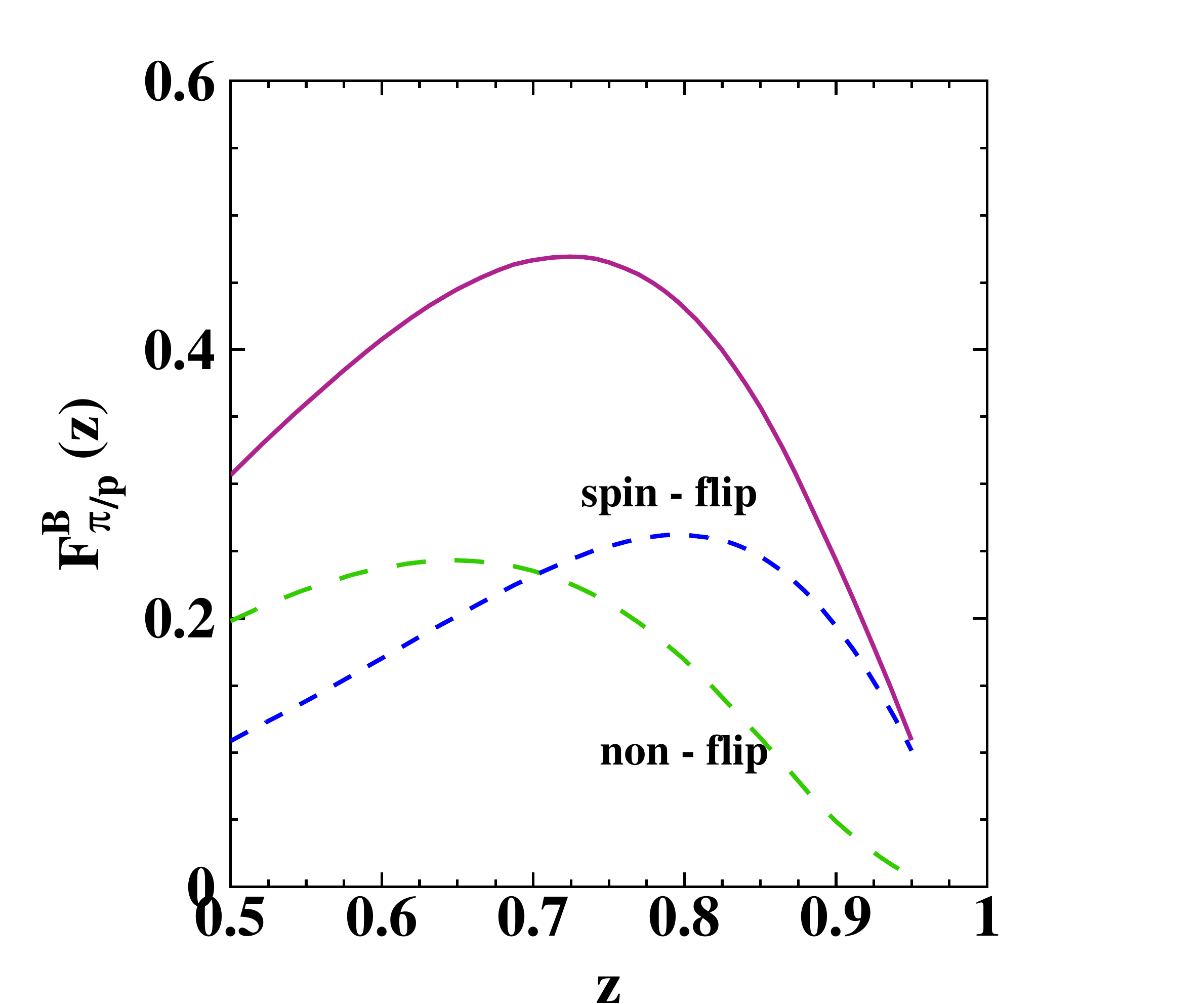}}}
\caption{\label{fig:F-B} (Color online) The $q$-integrated pion flux, calculated in the Born approximation,  vs neutron fractional momentum $z$.
Correspondingly the pion flux carries fraction $1-z$. The fluxes of pions flipping or non-flipping the nucleon helicity are depicted by dashed curves, while the solid curve presents the full pion flux. }
 \end{figure}
%%%%%%%%%%%%%%%%%%%%%

To enhance the statistics more, one can make use of all registered neutrons to extract the $\pi\pi$ total cross section,
\beq
\sigma(pp\to nXn)\bigr|_{z_{1,2}>z_{min}}=
\Phi^B(\tau)\,
\sigma^{\pi^+\pi^+}_{tot}(\tau s),\label{205}
\eeq
where the $pp\to nXn$ cross section is integrated over $z_{1,2}>z_{min}$.
The fractional coefficient $\Phi(\tau)$, within the Born approximation, is given by,
\beqn
\Phi^B(\tau)=
\int\limits_{z_{min}}^{z_{max}} \frac{dz_1}{1-z_1}\,
F^B_{\pi^+/p}(z_1)\,
F^B_{\pi^+/p}(z_2),
\label{210}
\eeqn
with 
 $z_2=1-\tau/(1-z_1)$.
The choice of $z_{min}$ defines the maximum value of $\tau\leq\tau_{max}=(1-z_{min})^2$. Further on we fix $z_{min}=0.5$. The upper integration limit is fixed by the relation $(1-z_{max})(1-z_{min})=\tau$. $\Phi^B(\tau)$  correlates with the amount of events detected in the interval $z_{min}<z<z_{max}$.
The interval of integration in (\ref{210}) shrinks to zero towards $\tau=\tau_{max}=0.25$, so the value of $\Phi^B(\tau)$ drops down. On the other hand, at small $\tau$ one of the rapidity gaps, shown in Fig.~\ref{fig:pi-pi}, become large (because $s$ is very large) and the pion exchange vanishes due to its low Regge intercept. Again, $\Phi^B(\tau)$
is falling due to Eq.~(\ref{210}).
Values of the coefficient $\Phi^B(\tau)$, calculated at $\sqrt{s}=7\TeV$  are plotted in Fig.~\ref{fig:PHI} vs $\tau$.
%%%%%%%%%%%%%%%%%%%%%
\begin{figure}[tbh]
\centerline{
  \scalebox{0.35}{\includegraphics{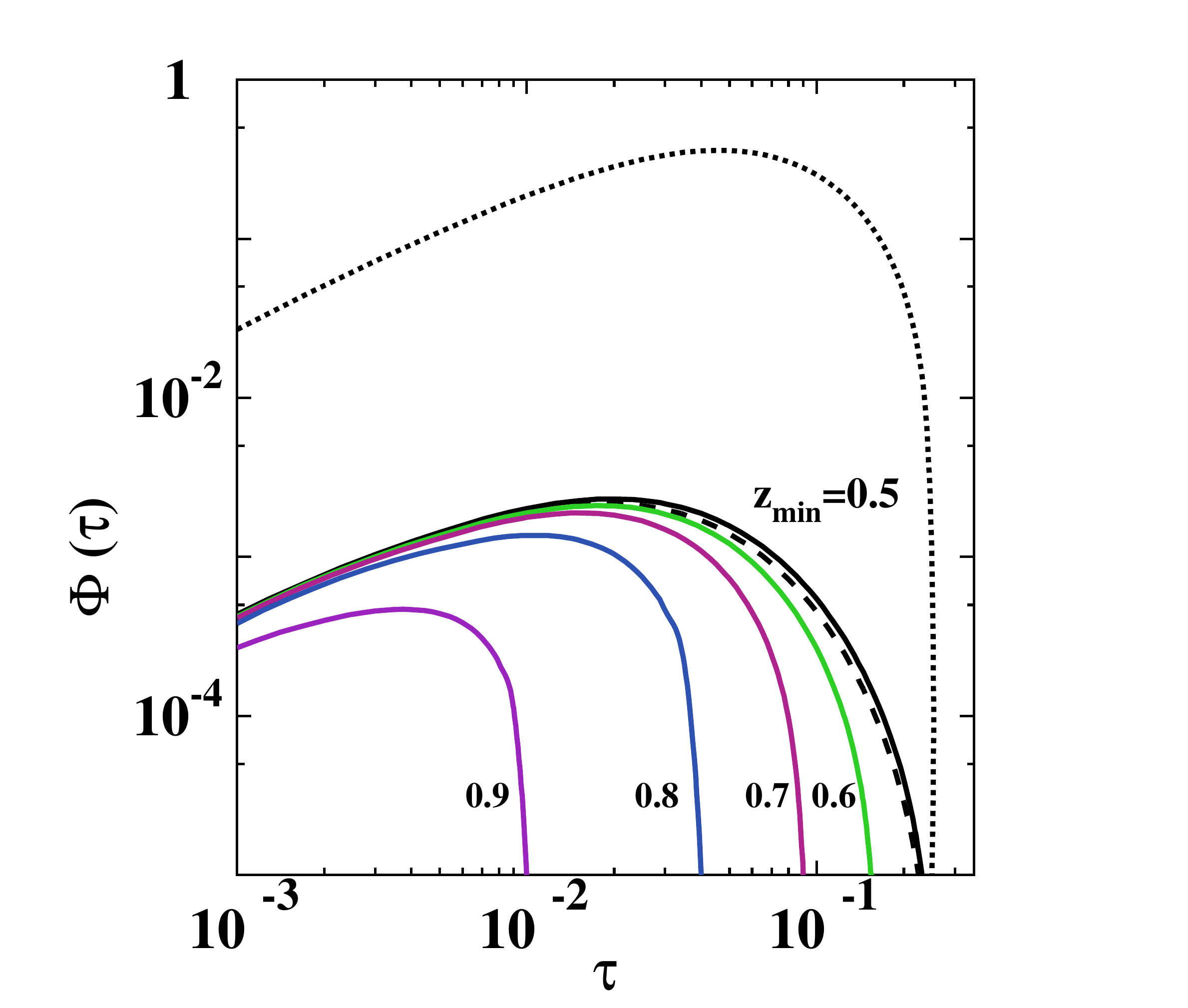}}}
\caption{\label{fig:PHI}  (Color online) The integrated flux of two pions at $z_{min}=0.5$ calculated in the Born approximation, Eq.~(\ref{197}), and with absorption corrections, Eq.~(\ref{850}), depicted by dotted and dashed curves respectively. The solid curves, including also the feed-down corrections, are calculated with $z_{min}=0.5-0.9$ (from right to left) as is marked on the plot.}
 \end{figure}
%%%%%%%%%%%%%%%%%%%%%
 
 Notice that a pion flux can originate from a transition
 $p\Rightarrow n+\pi$ with or without spin-flip of the nucleon helicity \cite{kpss,kpss-AN}.
 Correspondingly, we identify two different types of pion fluxes. 
 At this point we switch to amplitudes, because the survival amplitude introduced in the next section is very different for spin-flip and non flip amplitudes \cite{kpss,kpss-AN}. The amplitude of the process (\ref{100}) reads,
 %\begin{widetext}
  \beqn
&&A^B_{pp\to nXn}(z_1,\vec q_1;z_2,\vec q_2)=
A_{\pi^+\pi^+\to X}(\tau s)
\nonumber\\&\times&
\left\{
\bar\chi_{n1}\left[\sigma_3\, q_{L1}+
\frac{1}{\sqrt{z_1}}\,
\vec\sigma\cdot\vec q_1\right]\chi_{p1}\right\}\psi^B_\pi(z_1,\vec q_1)
\nonumber\\ &\times&
\left\{
\bar\chi_{n2}\left[\sigma_3\, q_{L2}+
\frac{1}{\sqrt{z_2}}\,
\vec\sigma\cdot\vec q_2\right]\chi_{p2}\right\}
\psi^B_\pi(z_2,\vec q_2),
\label{200}
 \eeqn
 % \end{widetext}
 where $\vec\sigma$ are Pauli matrices;  $\chi_{p,n}$ are the proton or
neutron spinors;  $\vec q_{1,2}$ are the transverse components of the neutrons momenta.

 At small $1-z\ll1$ the pseudo-scalar amplitudes $\psi^B_\pi$ have
the
triple-Regge form,
 \beq
\psi^B_\pi(z,\vec q)=\frac{\alpha_\pi^\prime}{8}\,
G_{\pi^+pn}(t)
\eta_\pi(t)
(1-z)^{-\alpha_\pi(t)}
\label{240}
 \eeq
 
 The amplitude of $\pi\pi$ collision is related to the total cross sections as,
 \beq
\sum\limits_X|A_{\pi^+\pi^+\to X}(\tau s)|^2 =
\tau s\,\sigma^{\pi^+\pi^+}_{tot}(\tau s)
\label{270}
 \eeq

Accordingly, the pion flux Eq.~(\ref{180}) can be split into two parts, corresponding to 
pion emission by the proton conserving or flipping its helicity,
 \beq
f^B_{\pi^+/p}(z,q)=f_{\pi^+/p}^{B(0)}(z,q)+f_{\pi^+/p}^{B(s)}(z,q),
\label{272}
\eeq
where
\beqn
f_{\pi^+/p}^{B(0)}(z,q) &=&
\frac{(1-z)}{z}\,q_L^2\,
\left|\psi^B_\pi(q,z)\right|^2
\label{274}\\
f_{\pi^+/p}^{B(s)}(z,q)&=&
-\frac{(1-z)}{z}\,(q_L^2+t)\,
\left|\psi^B_\pi(q,z)\right|^2\,.
\label{276}
 \eeqn
The corresponding $q$-integrated fluxes, Eq.~(\ref{197}), $F_{\pi^+/p}^{B(0)}(z)$
and $F_{\pi^+/p}^{B(s)}(z)$ are plotted by dashed curves in Fig.~\ref{fig:F-B}.

 \section{Absorptive corrections} \label{absorption}

The initial/final state inelastic interactions  lead to multi-particle production, which will fill the gaps, i.e. essentially reduce the fractional momenta, either $z_1$, or $z_2$, or both.
The no-interaction probability, usually called gap survival probability, certainly reduces the cross section compared with Eq.~(\ref{160}).

The absorptive corrections to the amplitude of a reaction are known to factorize in impact parameter representation. Therefore, we Fourier transform Eq.~(\ref{200}) to impact parameter space, introduce absorptive factors, and  transform the amplitude back to the momentum representation \cite{kpss,kpps-dis}.
\\

\subsection{The amplitude in impact parameters}\label{impact}

In the rest frame of of one of the colliding protons ($p_2$) the reaction Eq.~(\ref{100}) can be seen as the interaction of the pion flux in the proton $p_1$ (the upper pion in Fig.~\ref{fig:pi-pi}), i.e. $\pi+p_2\to X+n_2$.
The spin structure of this amplitude is given by the second factor in curly brackets in Eq.~(\ref{200}) and all the factors having subscript $2$ in (\ref{220}). A Fourier transform
of this part of the amplitude, $\int d^2q_2\exp(i\vec q_2\cdot\vec b_2)$, results in an amplitude which depends on the relative impact parameter $b_2$ between the colliding pion and $p_2$. Symmetrically, in the rest frame of the proton $p_1$, we obtain an amplitude dependent on the impact parameter $\vec b_1$ between $p_1$ and and the bottom pion in Fig.~\ref{fig:pi-pi}. Thus, making a double Fourier transformation we arrive at,
\begin{widetext}
\beqn
&&A^B_{pp\to nXn}(\vec b_1,z_1;\vec b_2,z_2)=
\int d^2q_2\,e^{i\vec q_2\cdot\vec b_2}
\int d^2q_1\,e^{i\vec q_1\cdot\vec b_1}
A^B_{pp\to nXn}(\vec q_1,z_1;\vec q_2,z_2) =
A_{\pi^+\pi^+\to X}(\tau s)
 \label{280}
 \\&\times&
 \left\{
\bar\chi_{n2}\left[\sigma_3\, q_{L2}\,\theta_B^{(0)}(b_2,z_2)-
i\,\frac{\vec\sigma\cdot\vec b_2}{\sqrt{z_2}\,b_2}\,
\theta_B^{(s)}(b_2,z_2)\right]\chi_{p2}\right\}
\left\{
\bar\chi_{n1}\left[\sigma_3\, q_{L1}\,\theta_B^{(0)}(b_1,z_1)-
i\,\frac{\vec\sigma\cdot\vec b_1}{\sqrt{z_1}\,b_1}\,
\theta_B^{(s)}(b_1,z_1)\right]\chi_{p1}\right\},
\nonumber
 \eeqn
\end{widetext}
where the partial amplitudes, spin non-flip and spin-flip, have similar structures \cite{kpss,kpss-AN,kpps-dis}, but depend on either $b_1,z_1$, or on $b_2,z_2$,
\beqn
\theta_B^{(0)}(b,z) &=& \int d^2q\,e^{i\vec b\vec q}\,
\psi^B_\pi(q,z)
\label{300}\\ &=&
\frac{\Omega_\pi(z)}{1-\beta_\pi^2\epsilon_\pi^2}\,
\left[K_0(\epsilon_\pi b)-K_0(b/\beta_\pi)\right]\,;
\nonumber
 \eeqn
  %%%%%%%%%%%%%%%
\beqn
\theta_B^{(s)}(b,z) &=& {1\over b}
\int d^2q\,e^{i\vec b\vec q}\,
(\vec b\cdot\vec q)\,\psi^B_\pi(q,z)
\label{320}
\\ &=&
\frac{\Omega_\pi(z)}{1-\beta_\pi^2\epsilon_\pi^2}\,
\left[\epsilon_\pi\,K_1(\epsilon_\pi b)-\frac{1}{\beta_\pi}\,K_1(b/\beta_\pi)\right].
\label{330}
\nonumber
 \eeqn
 Here
  \beq
\Omega_\pi(z) =\frac{1}{2}\,g_{\pi^+pn}\,
z(1-z)^{\alpha^\prime_\pi(m_\pi^2+q_L^2)}
e^{-R_\pi^2 q_L^2},
\label{340}
\eeq
contains the $q$-independent part of the flux Eq.~(\ref{180});
\beqn
\epsilon_\pi^2&=& z(q_L^2+m_\pi^2)\,,
\nonumber\\
\beta_\pi^2&=&{1\over z}\,\left[
R_\pi^2-\alpha_\pi^\prime\,\ln(1-z)\right].
\label{360}
 \eeqn

\section{Rapidity gap survival amplitudes}\label{survival}

The process (\ref{100}) results in the production of three colorless objects, $n_1$, $n_2$ and $X$.
Correspondingly, the overall survival amplitude contains three absorptive suppression factors,
\beqn
&&S_{pp\to nXn}(\vec b_1,z_1;\vec b_2, z_2)=
S^{\pi N}_{abs}\!\left(b_1,z_1,\frac{s_0}{1-z_1}\right)
\nonumber\\ &\times&
S^{\pi N}_{abs}\!\left(b_2,z_2,\frac{s_0}{1-z_2}\right)
S_{abs}^{NN}(b_{NN},s).
\label{380}
\eeqn
Here $S^{\pi N}_{abs}$ are the survival amplitudes for the rapidity gaps between the produced system $X$ and neutrons $n_1$ and $n_2$ respectively. The invariant mass  squared of the on-mass-shell fluctuations $p\to n\pi^+$
are $s_{1,2}=s_0/(1-z_{1,2})$, where $s_0\sim1/R_\pi^2$ is the mean value $q^2$.
In the survival amplitudes $S$ only possible inelastic interactions of the system $X$
with one of the two nucleons are included. The inelastic interaction between the 
spectator nucleons is excluded from $S$ in order to avoid a double counting, and is presented in Eq.~(\ref{380}) by a separate factor $S_{abs}^{NN}$, which depends on the impact parameter $b_{NN}$ of the $pp$ collision.

\subsection{Final-state interactions of the  system \boldmath$X$}\label{S1,2}

The inelastic $\pi+\pi\to X$ collision occurs at very high energy $\tau s$, and is a result of color gluonic exchange, leading to production of two color octet $\bar qq$ pairs, which are the debris of the colliding pions, as is illustrated in Fig.~\ref{fig:pi-p} (compare with Fig.~5 in \cite{kpss}). 
%%%%%%%%%%%%%%%%%%%%%
\begin{figure}[htb]
\centerline{
  \scalebox{0.3}{\includegraphics{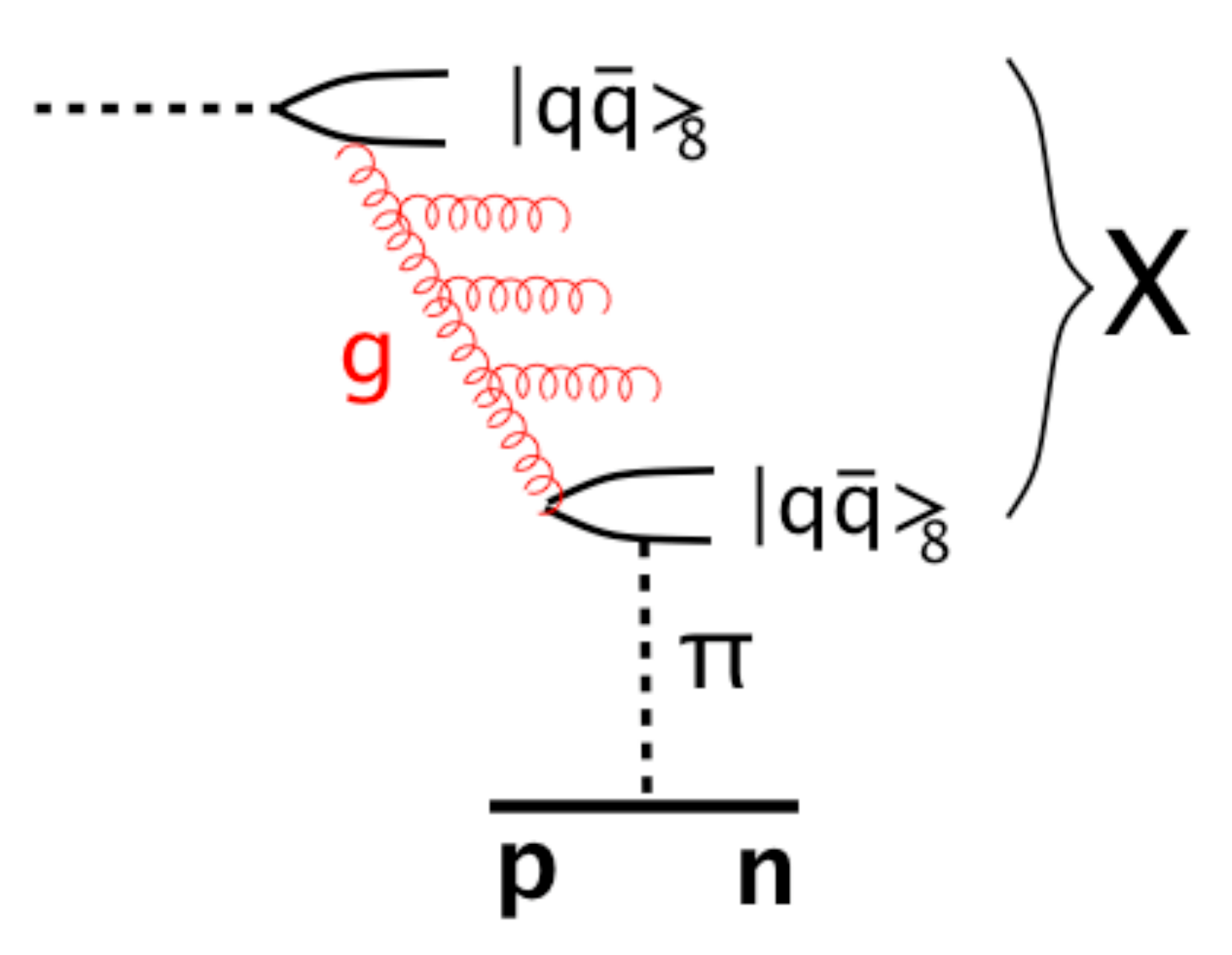}}}
\caption{\label{fig:pi-p} (Color online) Production of a color octet-octet dipole
in $\pi$-$\pi$ color-exchange collision in reaction $\pi^++p\to X+n$.}
 \end{figure}
%%%%%%%%%%%%%%%%%%%%%
Here we deal with absorptive corrections for leading neutron production in reaction $\pi p\to Xn$, which is natural to compare with the reaction $pp\to Xn$ studied in \cite{kpss}. Correspondingly, 
the absorption suppression factor $S_{4q}(b)$, caused by
possible inelastic interactions of the $\{\bar qq\}_8$-$\{\bar qq\}_8$ color octet-octet dipole with the nucleon, can be evaluated in close analogy with $S_{5q}(b)$, calculated in \cite{kpss}. One should only replace $B_{el}^{\pi p}\Rightarrow B_{el}^{\pi\pi}$ in the dipole size distribution, given by Eq.~(27)  in \cite{kpss}, and in the gap survival amplitude (28). We estimated  $B_{el}^{\pi\pi}(\tau s)$ relying on the known slope $B_{el}^{pp}(s)$ \cite{kpp2012} and assuming that $B_{el}^{pp}-B_{el}^{\pi\pi}\approx 4\GeV^{-2}$.
Notice that the dipole-nucleon amplitude depends on impact parameter $b$, which is the transverse distance between the center of gravity of the dipole and the nucleon. 

Further details of the calculations can be found in \cite{kpss,kpps-dis}.  
The result for the absorptive correction factor $S_{4q}(b)$, calculated at  $\sqrt{s}=7\TeV$ and $z_1=z_2=0.7$, is shown in Fig.~\ref{fig:S4q} by the dashed curve.
%%%%%%%%%%%%%%%%%%%%%
  \begin{figure}[htb]
\centerline{
  \scalebox{0.3}{\includegraphics{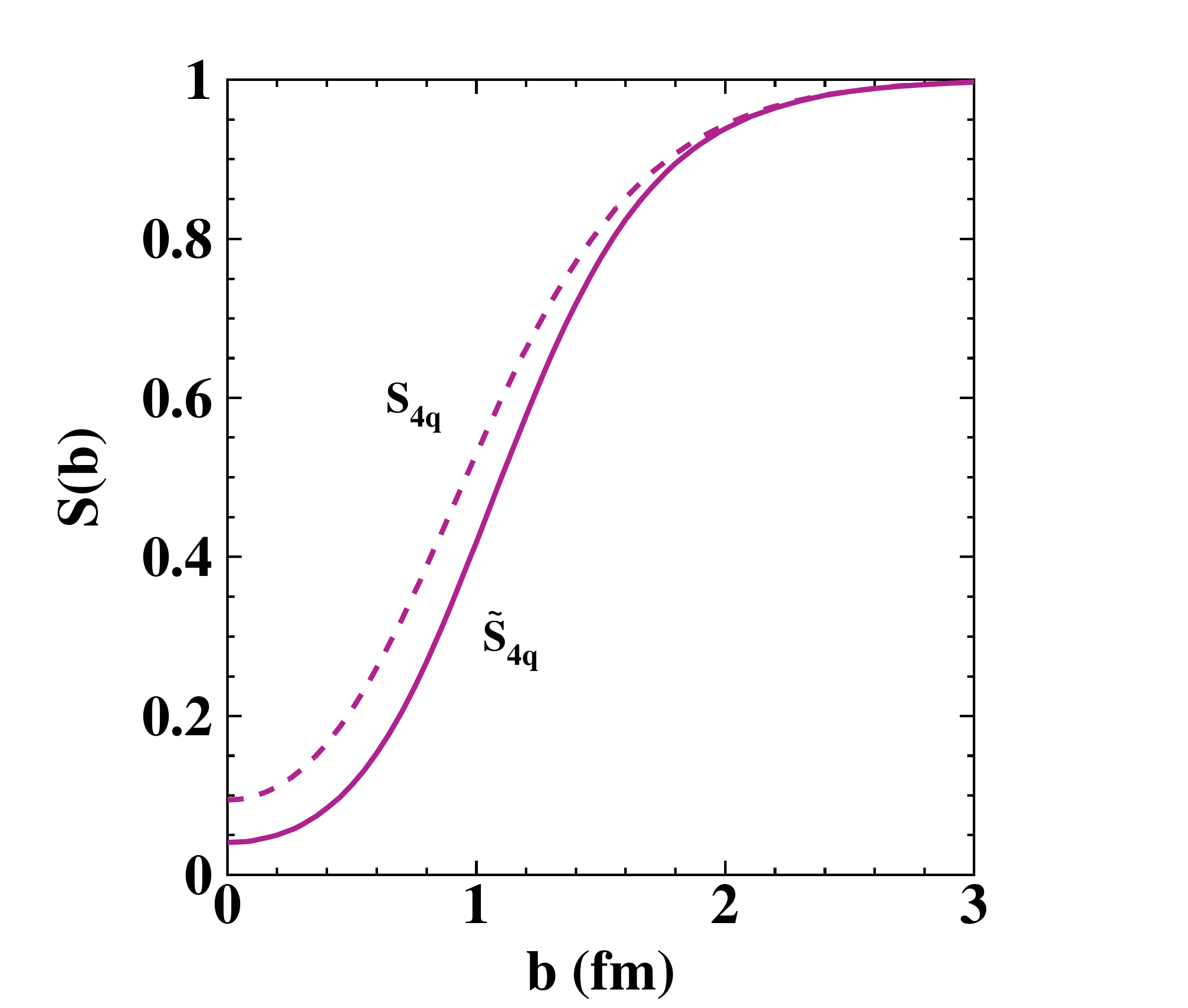}}}
 \caption{(Color online) Partial survival amplitudes including ($\tilde S_{4q}$) and excluding
($S_{4q}$)  interactions with the radiated gluons, as function of $b$, at collision energy $\sqrt{s}=7\TeV$ and $z_1=z_2=0.7$.}
 \label{fig:S4q}
 \end{figure}
 %%%%%%%%%%%%%%%%%%%%%

Apparently, the radiated gluons, depicted in Fig.~\ref{fig:pi-p}, also can interact and enhance the absorption corrections. 
The corresponding modification of the suppression factor $S_{4q} \Rightarrow \tilde S_{4q}$ was evaluated in \cite{kpps-dis},
\beq
\tilde S_{4q}(b) =
S_{4q}(b)e^{-\la n_g\ra\la f_{\bar qq}(b)\ra}.
\label{490}
\eeq
Here the set of radiated gluons was replaced by $\bar qq$ dipoles, with a size corresponding to the transverse glue-glue separation $r_{gg}$.
The mean value of the elastic partial amplitude of a $\bar qq$ dipole colliding with a proton at impact parameter $\vec b$, $\bigl\la f_{\bar qq}(\vec b,\vec r_{gg},s_{gp})\bigr\ra$, was calculated in \cite{kpps-dis}. Here we evaluate it at the mean gluon-proton energy, $\la s_{gp}\ra=(1-z)s/\ln(\tau s/s_0)$.

The important parameter, controlling the dipole size distribution, is the mean glue-glue separation, $r_0=0.3\fm$, which was adjusted in \cite{kst2} to reproduce the small triple-Pomeron coupling.

The mean number of radiated gluons $\la n_g\ra$ in Eq.~(\ref{490}) 
was evaluated for a soft Pomeron
in \cite{k3p} at,
\beq
\la n_g\ra=\frac{4\alpha_s}{3\pi}\,\ln\left(\frac{\tau s}{\tilde s_0}\right)
\label{488}
\eeq
where $\tilde s_0=30\GeV^2$. The QCD coupling at the semi-hard scale $Q^2\sim1/r_0^2$ was
estimated in \cite{k3p} within different models, which arrive at a similar value of $\alpha_s\approx 0.4$.
Notice that with this set of parameters the total and elastic $pp$ cross sections at LHC \cite{totem} were  predicted in \cite{k3p} with amazing accuracy \cite{kpp2012}.

The gluon absorption effect can be seen from comparison of $\tilde S_{4q}$ with $S_{4q}$ presented in Fig.~\ref{fig:S4q}.  
Although the density of radiated gluons Eq.~(\ref{488}) for a soft Pomeron is smaller than in DIS \cite{kpp2012}, the dipole amplitude in (\ref{490}) steeply rises with energy, and overcompensates the smallness of $\la n_g\ra$ at the high energy of LHC. Nevertheless,
 the effect of gluon absorption is significant only at small $b$. 
Notice that the $z$- and $s$-dependences of $\tilde S_{4q}(b)$ are rather weak within the energy range of the LHC.

As far as the absorption suppression factor, $S^{\pi N}_{abs}(b)=\tilde S_{4q}(b)$ is known, 
we are in a position to calculate the   amplitude of the process, corrected for absorption.
Let us forget for a moment the third factor $S_{abs}^{NN}(b_{NN})$ in Eq.~(\ref{380}).
Then, introducing the absorption factors into the partial amplitudes (\ref{300}) and (\ref{320})
and  Fourier transforming them back to momentum representation, we arrive at the amplitude in the same form as it was in Born approximation, Eq.~(\ref{200}), but with the new functions $\psi_\pi(z,q)$.
Their dependence on $z$ and $\vec q$, which are either $z_1,\vec q_1$, or $z_2,\vec q_2$,
has the form,
\beqn
&&\psi^{(0)}_\pi(z,q)=\frac{\Omega_\pi(z)}{2\pi(1-\beta_\pi^2\epsilon_\pi^2)}
\int\limits_0^\infty db\,b\,J_0(bq)
\nonumber\\ &\times&
\left[K_0(\epsilon_\pi b)-K_0\left({b\over\beta_\pi}\right)\right]
S^{\pi N}_{abs}(b,z);
\label{480}
 \eeqn

 \beqn
&&q\,\psi^{(s)}_\pi(z,q)=\frac{\Omega_\pi(z)}{2\pi(1-\beta_\pi^2\epsilon_\pi^2)}
\int\limits_0^\infty db\,b\,J_1(bq)\,
\nonumber\\ &\times&
\left[\epsilon_\pi\, K_1(\epsilon_\pi b)-
{1\over\beta_\pi}\,K_1\left({b\over\beta_\pi}\right)\right]
S^{\pi N}_{abs}(b,z).
\label{500}
 \eeqn

Thus, without the factor $S_{abs}^{NN}(b_{NN})$ in Eq.~(\ref{380}), including only the absorption corrections $S^{\pi N}_{abs}(b_{1,2})$, we recover the  factorized form of the cross section, Eq.~(\ref{160}), although 
with pion fluxes considerably modified by absorption.
Such a flux, $f_{\pi^+/p}$, contains two terms corresponding to pion emission by the proton, $p\to n\pi$, either preserving or flipping its helicity,
\beq
f_{\pi^+/p}(z,q)=f_{\pi^+/p}^{(0)}(z,q)+f_{\pi^+/p}^{(s)}(z,q),
\label{520}
\eeq
where
\beqn
f_{\pi^+/p}^{(0)}(z,q) &=&
\frac{(1-z)}{z}\,q_L^2\,
\left|\psi^{(0)}_\pi(q,z)\right|^2
\label{540}\\
f_{\pi^+/p}^{(s)}(z,q)&=&
\frac{(1-z)}{z^2}\,q^2\,
\left|\psi^{(s)}_\pi(q,z)\right|^2\,.
\label{560}
 \eeqn
We show the effect of absorption for $f_{\pi^+/p}^{(0)}(z,q)$ at $q=0$  in Fig.~\ref{fig:f0}. The absorption corrected flux (solid curve) turns out to be quite suppressed compared with the Born approximation (dashed curve).

The absorption corrected and $q$-integrated flux consists of spin non-flip and flip terms,
\beq
F_{\pi^+/p}(z_{1,2})=F^{(0)}_{\pi^+/p}(z_{1,2})+F^{(s)}_{\pi^+/p}(z_{1,2}),
\label{800}
\eeq
which have the form,
\beqn
F^{(0)}_{\pi^+/p}(z)&=&
\frac{(1-z)^3 m_N^2
\Omega^2_\pi(z)}{4\pi^3 z^2(1-\beta_\pi^2\,\epsilon_\pi^2)^2}
\int d^2b\,\left[S^{\pi N}_{abs}(b,z)\right]^2
\nonumber\\ &\times&
\left[K_0(\epsilon_\pi b)-K_0\left({b\over\beta_\pi}\right)\right]^2;
\label{820}
 \eeqn
and
 \beqn
F^{(s)}_{\pi^+/p}(z)&=&
\frac{(1-z)
\Omega^2_\pi(z)}{4\pi^3 z^2(1-\beta_\pi^2\,\epsilon_\pi^2)^2}
\int d^2b\,\left[S^{\pi N}_{abs}(b,z)\right]^2
\nonumber\\ &\times&
\left[\epsilon_\pi\, K_1(\epsilon_\pi b)-
{1\over\beta_\pi}\,K_1\left({b\over\beta_\pi}\right)\right]^2.
\label{840}
 \eeqn
 
In Fig.~{\ref{fig:F(z)} we present the full effective flux of pions, corrected for the absorption factor $S^{\pi N}_{abs}(b)$ and integrated over $q$, as well as its spin-flip and non-flip components.

%%%%%%%%%%%%%%%%%%%%%
  \begin{figure}[htb]
\centerline{
  \scalebox{0.3}{\includegraphics{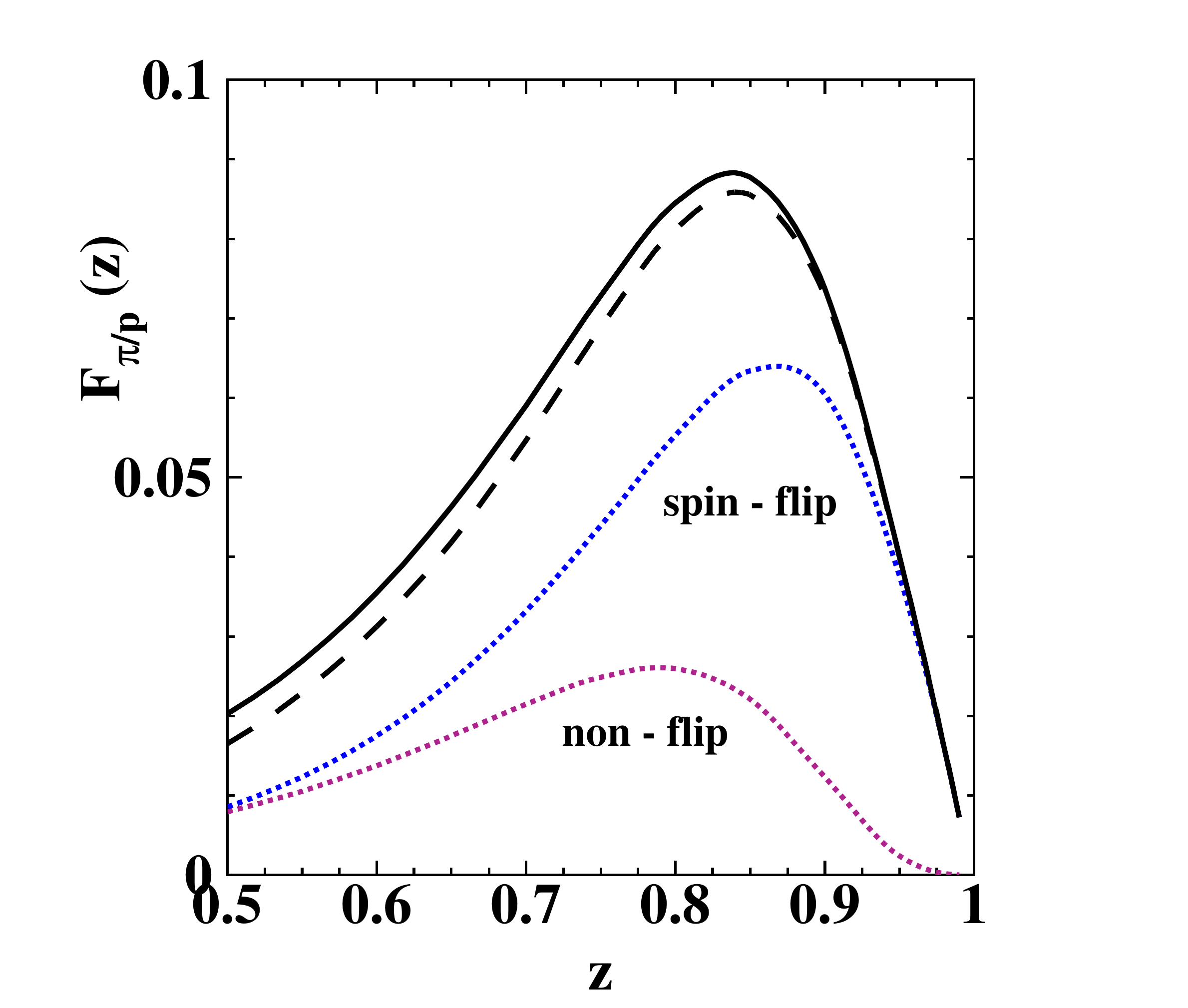}}}
 \caption{(Color online) The $q$-integrated pion flux $F_{\pi^+/p}(z)$ calculated with Eqs.~(\ref{800})-(\ref{840}), which include absorption. The dotted and dashed curves show the spin parts and the full flux respectively. The full flux enhanced by the feed-down corrections is depicted by solid curve. 
}
 \label{fig:F(z)}
 \end{figure}
 %%%%%%%%%%%%%%%%%%%%%
 
 Notice that the $q$-integrated pion flux $F_{\pi^+/p}(z)$ includes the contribution of large $q$, which it the kinematic region rather far from the pion pole. So, the correctness
 of extrapolation, the pion-nucleon formfactor, and the pion dominance become questionable.
 Therefore one should try to restrict the range of integration, $q < q_{max}$ to exclude large $q$. As usual, this is subject to a compromise between the desirable smallness of $q_{max}$ and statistics. Besides, the available ZDCs have a limited range of $q$ anyway.
To demonstrate the variation of  $F_{\pi^+/p}(z)$ as function of $q_{max}$ we performed calculations with different $q_{max}=0.1-0.4\GeV$ as is marked in Fig.~\ref{fig:F-qmax}.
 %%%%%%%%%%%%%%%%%%%%%
  \begin{figure}[htb]
\centerline{
  \scalebox{0.3}{\includegraphics{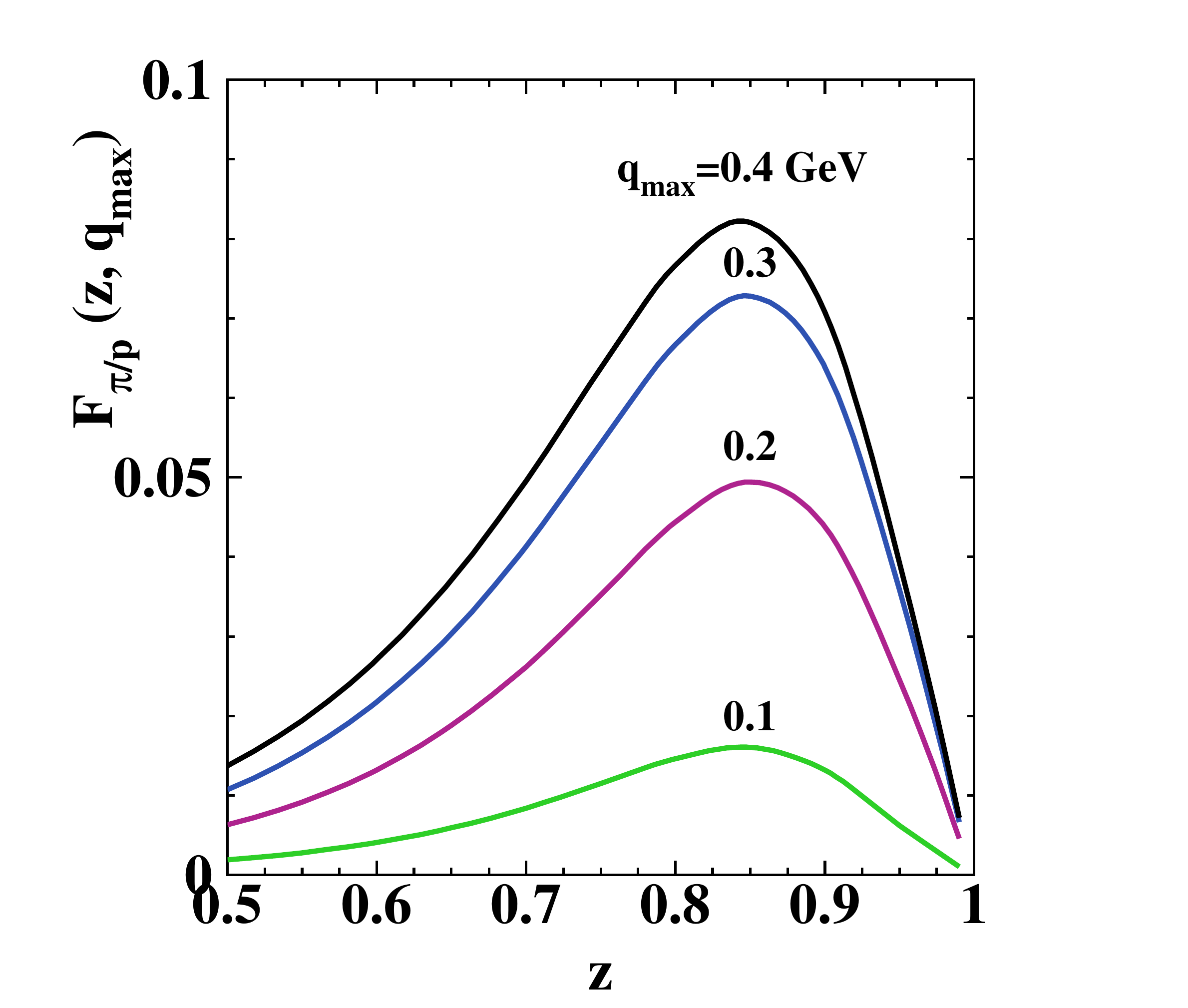}}}
 \caption{(Color online) The same as in Fig.~\ref{fig:F(z)}, but integrated over $q$ up to $q_{max}=0.1-0.4\GeV$, as is marked on the plot. The effects of absorption and the feed-down corrections are included. 
}
 \label{fig:F-qmax}
 \end{figure}
 %%%%%%%%%%%%%%%%%%%%%

\subsection{Feed-down corrections} \label{feed-down}

The absorption suppression factors reject any possibility of inelastic interactions between the participating partons,
assuming that it would lead to multi-particle production and termination of the rapidity gap, related to the pion exchange. This is why this suppression is also called gap survival probability.
However, a part of the inelastic cross section is related to other rapidity-gap processes, e.g. diffraction. A nucleon experienced such an interaction
re-appears with a reduced
fractional momentum $z'<1$ and still can
contribute to the process under consideration. 
%The result of such multiple initial or final state interactions of the participating nucleons can be treated as energy loss. 
This effect named "migration"  in \cite{kkmr}, was found to produce a sizeable excess of neutrons at medium-large values of $z$. 
We, however, calculate these corrections differently and arrive at a smaller effect. First of all, the momentum distribution of the produced nucleons was calculated in \cite{kkmr} according to the Kancheli-Mueller graph, which is reasonable only at small $z'\ll 1$,
while we are interested in $z>0.5$, i.e. $z'\gsim 0.7$, because  according to Fig.~\ref{fig:F-B} the $p\to n$ transition peaks at $z\approx 0.7$-$0.8$. At large $z'$ the triple-Regge description is more appropriate, and the diffractive term, which steeply rises at $z'\to1$, essentially reduces energy loss, i.e. the shift in $z$ (migration).

Another drawback of the calculations performed in \cite{kkmr}, was the probabilistic treatment of multiple interactions. The energy loss was calculated in the Bethe-Heitler regime,
i.e. assuming that the full spectrum of particles is produced in each re-scattering. However, coherence, or the Landau-Pomeranchik effect, is known to reduce significantly the rate of energy loss \cite{bdmps}. In what follows we rely on a fit to the rapidity-gap cross sections, so multiple interactions are included by default.
 
Treating energy loss in terms of the Fock state decomposition, one can say that the higher components of the proton, contain besides the pion also a flux of Reggeons $\mathbb{R}$ ($f,\ \omega$), or Pomeron, $\mathbb{P}$, etc. In those fluctuations, which are released to mass shell by the interaction, energy sharing leads to a reduction of the neutron momentum, i.e. to a feed-down of the smaller $z$-regions (migration).

Keeping the feed-down corrections in  the lowest order, the pion flux can be modified as, 
%%%%%%%%%%%%%
\beqn
f_{\pi/p}(z,q)&\Rightarrow&f_{\pi/p}(z,q)+\Delta f_{\pi/p}(z,q);
\label{150a}\\
F_{\pi/p}(z)&\Rightarrow&F_{\pi/p}(z)+\Delta F_{\pi/p}(z).
\label{150b}
\eeqn

Let us start with the Pomeron contribution to $\Delta f_{\pi/p}$. 
While the uncorrected pion flux is related to the collision, $\pi\pi\to X$ with $M_X^2=\tau s$,
the correction includes a double collision: (i) $\mathbb{P(R)}\pi\to X^{\prime}$ with $M_{X^{\prime}}^2\equiv s^{\prime}$; and (ii) $\pi\pi\to X^{\prime\prime}$ with $M_{X^{\prime\prime}}^2\equiv s^{\prime\prime}$, as is illustrated in Fig.~\ref{fig:feeddown}.
%%%%%%%%%%%%%%%%%%%%%
\begin{figure}[htb]
\centerline{
  \scalebox{0.4}{\includegraphics{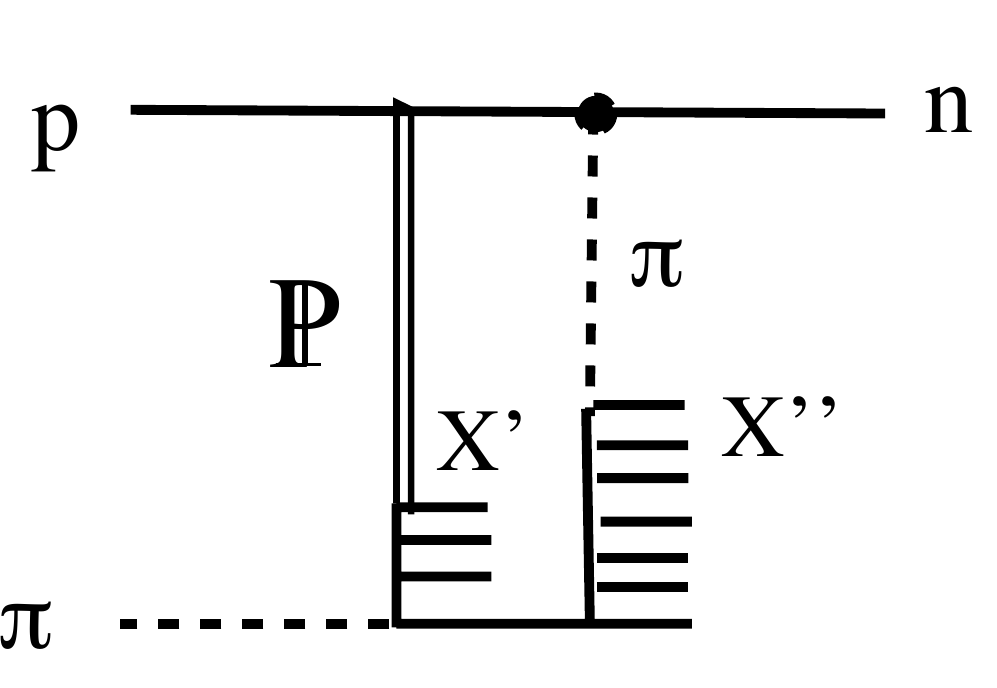}}}
\caption{\label{fig:feeddown} Graphical illustration for the double-step correction to neutron
production.}
 \end{figure}
%%%%%%%%%%%%%%%%%%%%%
We treat the interaction of the pair flux, \{$\mathbb{P}$+$\pi$\}, from one of the colliding protons,  with the pion flux originated from another proton, in a Glauber-like way. The amplitude is a sum of two terms, the $\pi\pi$ one, corresponding to (\ref{520}); and the double-scattering term, which has positive sign, differently from the shadowing term in the Glauber formula, in accordance with the AGK cutting rules.
Interaction with only the Pomeron flux does not lead to neutron production, so is excluded.

Since we employ the phenomenological cross section for the Pomeron exchange in Fig.~{fig:feeddown}, it includes all possible rescattering corrections.
What is missing, however, is a possibility of
of nucleon excitation in the intermediate state between the pion and multi-Pomeron exchanges, as is illustrated in Fig.~\ref{fig:pom-pi}. 
%%%%%%%%%%%%%%%%%%%%%
\begin{figure}[htb]
\centerline{
  \scalebox{0.3}{\includegraphics{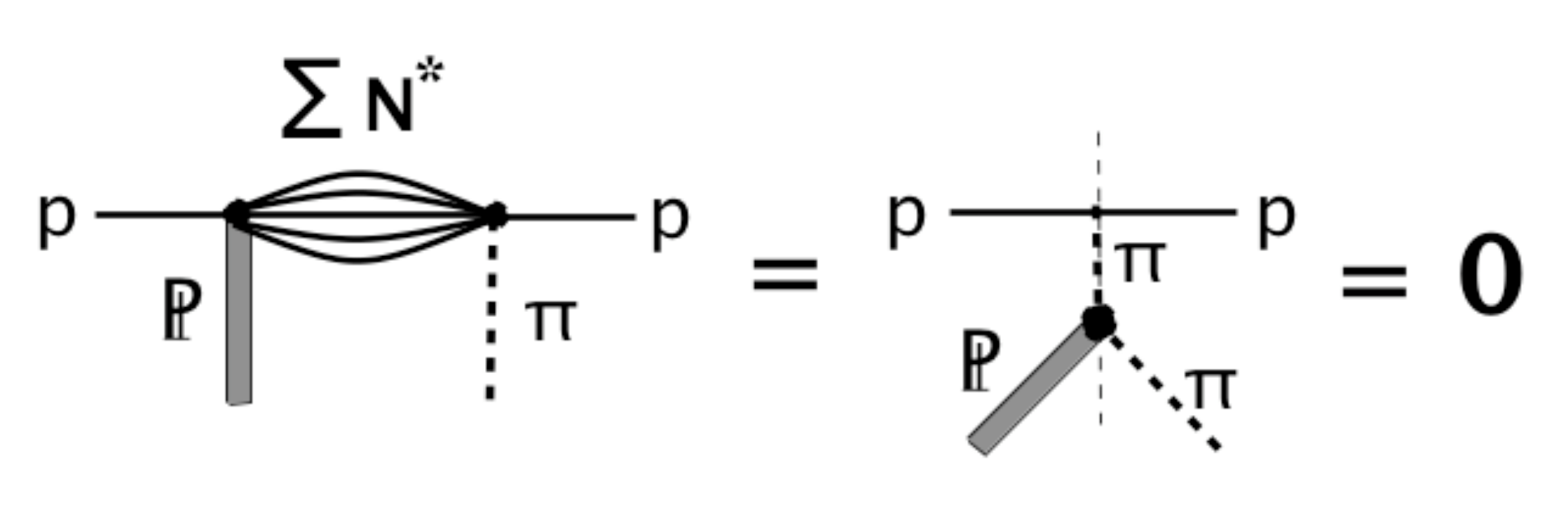}}}
\caption{\label{fig:pom-pi} Correction to the eikonal approximation due to nucleon excitations in the intermediate state between the pion-exchange and absorption amplitudes.}
 \end{figure}
%%%%%%%%%%%%%%%%%%%%%
This correction has not been discussed so far in previous calculations of absorptive corrections.
It includes a non-trivial overlap of the final state in proton excitation, either diffractive, or
via pion exchange. Apparently, many states are excluded (large mass diffraction, $\Delta$ excitations, etc.), moreover, we found this correction zero, or at least much suppressed.
Indeed, relying of the concept of duality, the sum of different $s$-channel excitation can be replaced by a $t$-channel Reggeons, as is shown in the middle picture Fig.~\ref{fig:pom-pi}. The Reggeon, which undergoes the unitarity cut, must be iso-vector with negative G-parity. Pion and its daughter Reggeons are real at $t=0$ and give no contribution. The axial-vector $a_1$ is an extremely weak singularity, which couplings in $pp$ collisions are suppressed by more than an order of magnitude compared with vector mesons \cite{marage}. Thus, we can safely neglect this kind of intermediate excitations.

As long as the corrections depicted in Fig.~\ref{fig:pom-pi} can be neglected, the amplitudes of the double-step process depicted in Fig.~\ref{fig:feeddown}  factorise in impact parameters, 
correspondingly in momentum representation the correction term in (\ref{150a}) reads,
\beqn
&&\Delta f_{\pi/p}(z,q) = \frac{1}{\sigma_{\pi\pi}(\tau s)}
\int\limits_{z}^{z^\prime_{max}}\!\!\frac{dz^\prime}{z^\prime}
\label{170a}\\ &\times&
 \left|\int \frac{d^2q^\prime}{(2\pi)^2}\,
A_{p\pi\to pX^\prime}(z^\prime,\vec q^{\,\prime})\,
A_{p\pi\to nX^{\prime\prime}}(z/z^\prime,\vec q-\vec q^{\,\prime})
\right|^2,
\nonumber
\eeqn
where ${z^\prime_{max}}={\rm min}\{z^\prime/z_{max},1\}$.

To make use of the triple-Regge phenomenology, we need to switch from convolution of amplitudes to cross sections. This is easy if to assume that the $q$-dependence of the amplitudes has a gaussian form, as is confirmed with a good accuracy by direct calculation of the $q$ dependence in \cite{kpps-dis} and by data \cite{zeus,h1}. Then the amplitude convolution, Eq.~(\ref{170a}), can be replaced by
%%%%%
\begin{widetext}
\beqn
\Delta f_{\pi/p}(z,\vec q)&=&
\sum\limits_{i,j,k=\mathbb{P,R}}
\int\limits_{z}^{z^\prime_{max}}\frac{dz^\prime}{(z^\prime)^2}\,
(1-z^\prime)^{\alpha_k(0)-\alpha_{i}(t_{min}^\prime)-\alpha_{j}(t_{min}^\prime)}
\left(\frac{\tilde s}{s_0}\right)^{\alpha_k(0)-1}
\frac{G^{\pi p}_{ijk}(t_{min}^\prime)}
{(B_{ijk}+B_{\pi})^2}
\nonumber\\ &\times&
\exp\left[-q^2\frac{B_{ijk}B_\pi}{B_{ijk}+B_{\pi}}\right]
f_{\pi/p}(z/z^\prime,q=0)\,
\frac{\sigma_{\pi\pi}(s^{\prime\prime})}
{\sigma_{\pi\pi}(\tau s)}.
\label{160a}
\eeqn
\end{widetext}
%%%%%
Here $s^\prime=(1-z^\prime)(1-z_2)s$; $s^{\prime\prime}=(1-z/z^\prime)(1-z_2)s$; 
$\tilde s=(1-z_2)s$;
$t^\prime$ is given by (\ref{140}) replacing $z\Rightarrow z^\prime$; $t_{min}^\prime=t^\prime(q=0)$.

The ratio of $\pi\pi$ cross sections in Eq.~(\ref{160a}) can be estimated assuming a power dependence $s^\epsilon$ with $\epsilon\approx 0.1$,
\beq
\frac{\sigma_{\pi\pi}(s^{\prime\prime})}{\sigma_{\pi\pi}(\tau s)}=
\left(\frac{1-z/z^\prime}{1-z}\right)^\epsilon.
\label{170d}
\eeq

The effective triple-Regge vertices $G_{ijk}(t)$ were fitted in \cite{kklp} to data for the rapidity-gap process $pp\to pX$, and the results are presented in Table~\ref{solution} of \ref{3R}.
This 40-years old fit predicts the recent data for diffractive inclusive production at the LHC amazingly well, as is demonstrated in \ref{3R}.

Assuming approximate Regge factorization (certainly broken by absorption), the effective triple-Regge coupling $G^{\pi p}_{ij\mathbb{P}}(t)$ in the process $p\pi\to pX^\prime$ can be related to $G^{pp}_{ij\mathbb{P}}(t)$,  known from data on $pp\to pX$,
$
G^{\pi p}_{ij\mathbb{P}}(t)=
(\sigma^{\pi p}_{tot}/\sigma^{pp}_{tot})\,
G^{pp}_{ij\mathbb{P}}.
$
The $\pi p$ total cross section at high energies is not known, but for our estimate it can be approximated by $\sigma^{\pi p}_{tot}/\sigma^{pp}_{tot}\approx2/3$. The same relation is natural to assume for the $ij\mathbb{R}$ vertex, because it is dominated by the $f$-Reggeon exchange ($\mathbb{R}\approx f$), while $\omega$ exchange is suppressed \cite{kklp}.

The triple-Regge $q^2$-slopes $B_{ijk}$ in (\ref{160a}) according to Eq.~(\ref{a100}) are given by,
\beq
B_{ijk}={1\over z^\prime}\,\left[R_{ijk}^2-(\alpha_i^\prime+\alpha_j^\prime)\ln(1-z^\prime)\right],
\label{180a}
\eeq
where $\alpha^\prime_i$ is the slope of the corresponding Regge trajectory, and the slope parameters $R^2_{ijk}$ fitted to $pp\to pX$ data are also presented in Table~\ref{solution} in \ref{3R}. The slope parameter $B_\pi$ describes the $q$-dependence of the pion flux 
$f_{\pi/p}(z_1,q)$, which we assumed above to have the gaussian form. It can be estimated as $B_\pi=f_{\pi/p}(z_1/z^\prime,q=0)/F_{\pi/p}(z_1/z^\prime)$.

Now we are in a position to perform numerical calculations for the feed-down corrections.
The corrected pion flux Eq.~(\ref{150a}) at $q=0$ is depicted by solid curve in Fig.~\ref{fig:f0}. The difference between the dashed and solid curve is hardly visible, so the feed-down correction is very small. On the other hand, the $q$-slope of the double step process is significantly smaller compared with $_\pi$, as one can see from Eq.~(\ref{160a}).
Therefore, 
the feed-down correction to the $q^2$-integrated flux, Eq.~(\ref{150b}), is much bigger, as one can see in Fig.~\ref{fig:F(z)}. The fluxes $F_{\pi/p}(z)$ including, or excluding the feed-down corrections  are plotted  in Fig.~\ref{fig:F(z)} by solid and dashed curves respectively. 
The correspondingly corrected combined flux of two pions, $\Phi(\tau)$, is shown by solid curve in Fig.~\ref{fig:PHI}.

\subsection{Absorption caused by the spectator nucleons} \label{S(NN)}

Although the factorized form of the cross section, Eq.~(\ref{160}) still holds for the fluxes suppressed by the absorption damping factors $S^{\pi N}_{abs}(b)$, 
the introduction of  the third factor $S_{abs}^{NN}(b_{NN})$ in Eq.~(\ref{380}) leads to a breakdown of such a factorized relation.

The absorption factor $S_{abs}^{NN}(b_{NN})$ presented by the grey strips in Fig.~\ref{fig:pi-pi-abs-k}, has the form \cite{kpss},
\beq
S_{abs}^{NN}(b_{NN},s))=1-\im\,f_{el}^{NN}(b_{NN},s),
\label{620}
\eeq
where $f_{el}^{NN}(b_{NN},s)$ is the partial elastic amplitude of $pp$ scattering.
If the amplitude is taken directly from data (see below), it includes all effects of unitarization, so can be used as is. What is missed, however, is the possibility of excitation of the nucleon between the multi-Pomeron and pion exchanges (see Fig.~\ref{fig:pi-pi-abs-k}). This correction was discussed above in section~\ref{feed-down} and found to be vanishingly small (see Fig.~\ref{fig:pom-pi}). 
 
 The impact parameter $b_{NN}$ is related to $b_1$ and $b_2$, which control the absorptive factors $S^{\pi N}_{abs}(b_{1,2})$
in Eq.~(\ref{380}), as $\vec b_{NN}=\vec b_1+\vec b_2$. 
Indeed, $b_1$ is the impact parameter between the center of gravity of the color octet-octet dipole, which is treated as a Fock component of the incoming pion $\pi_1$ (see Fig.~\ref{fig:pi-pi}), relative to the proton target. And vice versa, $b_2$ is the transverse distance between the same center of the dipole and another colliding proton.

For further calculations is convenient to switch to momentum representation,
\beqn
&&S_{abs}^{NN}(\vec b_1+\vec b_2)=
\int d^2b_{NN}\,S_{abs}^{NN}(\vec b_{NN})\,\delta(\vec b_{NN}-\vec b_1-\vec b_2)
\nonumber \\ &=&
\frac{1}{(2\pi)^2}\int d^2k\, d^2b_{NN}\,S_{abs}^{NN}(\vec b_{NN})\exp\left[i\vec k(\vec b_{NN}-\vec b_1-\vec b_2)\right]
\nonumber \\ &=&
\int d^2k\, S_{abs}^{NN}(\vec k)\exp\left[-i\vec k(\vec b_1+\vec b_2)\right]
\label{580}
\eeqn
The two new factors from Eq.~(\ref{580}), $e^{-i\vec k\vec b_1}$ and $e^{-i\vec k\vec b_2}$, should be included into Eqs.~(\ref{480}) and (\ref{500}) in the integration over $b_1$ and $b_2$ respectively. The final expression for the absorption corrected cross section reads,
 %%%%%%%%%%%%
\begin{widetext}
 \beqn
\frac{d\sigma^B(pp\to nXn)}{dz_1dz_2\,dq_{1}^2dq_{2}^2}
&=& 2\int d^2k\,d^2k'\,
\im f_{el}^{\pi\pi}(\vec k+\vec k',\tau s)\,
S_{abs}^{NN}(\vec k)\,S_{abs}^{NN}(\vec k')
\nonumber\\ &\times&
\frac{1-z_1}{z_1^2}\left[z_1 q_{L1}^2\,\psi_0(z_1,\vec q_1+\vec k)\psi_0^*(z_1,\vec q_1+\vec k')+
q_1^2\,\psi_s(z_1,\vec q_1+\vec k)\psi_s^*(z_1,\vec q_1+\vec k')
\right]
\nonumber\\ &\times&
\frac{1-z_2}{z_2^2}\left[z_2 q_{L2}^2\,\psi_0(z_2,\vec q_2+\vec k)\psi_0^*(z_2,\vec q_2+\vec k')+
q_2^2\,\psi_s(z_2,\vec q_2+\vec k)\psi_s^*(z_2,\vec q_2+\vec k')
\right]
\label{600}
\eeqn
\end{widetext}
This integral is presented graphically in Fig.~\ref{fig:pi-pi-abs-k}.
%%%%%%%%%%%%%%%%%%%%%
\begin{figure}[htb]
\centerline{
  \scalebox{0.35}{\includegraphics{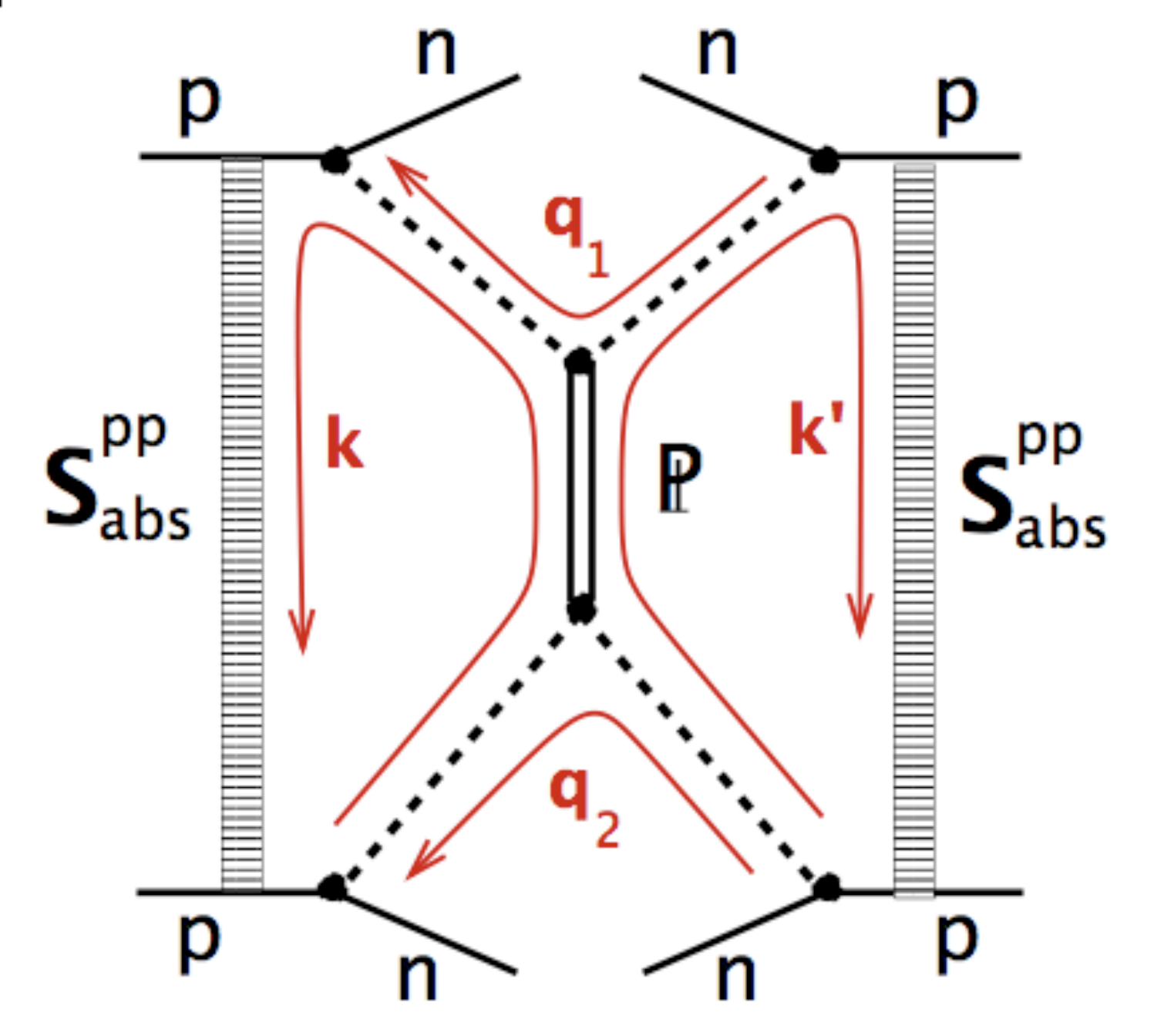}}}
\caption{\label{fig:pi-pi-abs-k} (Color online) Graphical representation for the absorption corrected cross section of reaction $pp\to nXn$. Dashed lines show the pions, the double line shows the Pomeron exchange. The flows of transverse momenta, $\vec k$, $\vec k'$, $\vec q_1$ and $\vec q_2$ are depicted by red lines. }
 \end{figure}
%%%%%%%%%%%%%%%%%%%%%

According to Eqs.~(\ref{620}) and (\ref{580}), in momentum representation
 \beq
S_{abs}^{NN}(k,s))=
\delta(k)-\im\,f_{el}^{NN}(k,s).
\label{640}
 \eeq
 Here and in what follows we do not discriminate between $pp$, $nn$ and $pn$ cross sections at the LHC energies, so label them as $NN$.
The elastic amplitude is related to the differential $NN$ cross section as,
\beq
\frac{d\sigma_{el}^{NN}}{dk^2}=
\frac{1+\rho^2(k)}{4\pi}
\left|\im\,f_{el}^{NN}(k,s)\right|^2,
\label{660}
\eeq
where $\rho(k)=\re f_{el}(k)/\im f_{el}(k)$ gives in (\ref{660}) a small correction of about $2\%$ (at small $k$), so can be neglected. The differential elastic cross section is known directly from data, and it includes by default all the effects of unitarization. These effects lead to a small deviation from the Gaussian $k$-dependence, i.e. a $k$-dependence of the slope $B(k)$, as is demonstrated in \cite{kpp2012}. Nevertheless the available rather precise data from the TOTEM experiment \cite{totem} agree also with a constant $B$.  So for the sake of simplicity we adopt the Gaussian parametrization from \cite{totem,alpha} for the elastic amplitude in (\ref{660}),
\beq
\im f_{el}^{NN}(k,s)={1\over2}\sigma^{NN}_{tot}(s)\,
e^{-k^2 B^{NN}_{el}(s)/2},
\label{680}
\eeq
where the slope measured at $\sqrt{s}=7\TeV$ is quite large,
$B_{el}^{NN}=20\GeV^{-2}$. Therefore the $k$-dependence of $S_{abs}^{NN}(k)$ is much steeper compared with the other terms in the square brackets under the integral in (\ref{600}), which comes mainly from the amplitude of elastic $\pi-\pi$ scattering,
\beqn
\im f_{el}^{\pi\pi}(\vec k\!+\!\vec k',\tau s)=
{1\over2}\sigma^{\pi\pi}_{tot}(\tau s)
e^{-(\vec k+\vec k')^2 B^{\pi\pi}_{el}(\tau s)/2}\!\!.
\label{700}
\eeqn
The $k$-dependence of the $\pi^+pn$ vertices is weak,  and the slope used in our calculations $R_\pi^2=0.3\GeV^{-2}$ is negligibly small compared to the elastic slopes of $NN$ and $\pi\pi$ amplitudes. Therefore, we can safely neglect it and fix $k=k'=0$ in all four $\pi^+pn$ vertices 
shown in Fig.~\ref{fig:pi-pi-abs-k}, i.e. in all expressions in square brackets in (\ref{600}).
Then we arrive at the final form of the cross section,
 \beqn
\!\!\!\! \frac{d\sigma(pp\to nXn)}{dz_1dz_2\,dq_{1}^2dq_{2}^2}&=&
f_{\pi^+/p}(z_1,q_{1})\,
\sigma^{\pi^+\pi^+}_{tot}(\tau s)
\nonumber\\&\times& 
f_{\pi^+/p}(z_2,q_{2})
D_{abs}^{NN}(s,z_1,z_2).
\label{720}
\eeqn
The absorption corrected effective pion fluxes, $f_{\pi^+/p}(z_{1,2},q_{1,2})$ are given
by Eqs.~(\ref{520})-(\ref{560}). The effect of interaction of the spectator nucleons 
is given by the last damping factor, 
\beq
D_{abs}^{NN}(s,z_1,z_2)=1-2I_1+I_2,
\label{730}
\eeq
where,
\beqn
I_1&=& \frac{2}{(2\pi)^2\sigma_{tot}^{\pi\pi}(\tau s)}
\int d^2k\,
\im f_{el}^{\pi\pi}(k,\tau s)\,
\im f_{el}^{NN}(k,s)
\nonumber\\ &=&
\frac{\sigma_{tot}^{NN}(s)/4\pi}
{B_{el}^{NN}(s) + B^{\pi\pi}_{el}(\tau s)};
\label{740}
\eeqn

\beqn
I_2 &=& 
\frac{2}{(2\pi)^4\sigma_{tot}^{\pi\pi}(\tau s)}
\int d^2k\,d^2k'\,
\im f_{el}^{\pi\pi}(\vec k+\vec k',\tau s)\,
\nonumber\\ &\times&
\im f_{el}^{NN}(k,s)
\im f_{el}^{NN}(k',s)
\nonumber\\ &=&
\frac{\left(\sigma_{tot}^{NN}/4\pi\right)^2}
{\left(B_{el}^{NN}(s)\right)^2+
2B_{el}^{NN}(s)B^{\pi\pi}_{el}(\tau s)}.
\label{760}
\eeqn
As an example, we can estimate the damping factor $D_{abs}^{NN}$ at $\sqrt{s}=7\TeV$
and the mean value of $\tau=0.5$. According to \cite{totem,alpha} $\sigma_{tot}^{NN}=98\mb$ and $B_{el}^{NN}=20\GeV^{-2}$. Relying on Regge factorization, we expect 
an energy independent difference $\Delta=B_{el}^{NN}-B_{el}^{\pi N}\approx 3\GeV^{-2}$. Correspondingly, the difference with the $\pi\pi$ slope is twice bigger, 
$B_{el}^{\pi\pi}=B_{el}^{NN}-2\Delta=14\GeV^{-2}$. With the standard energy dependence of elastic slopes, given by the term $2\alpha_{\Pom}^\prime\ln(s/s_0)$, with $\alpha_{\Pom}^\prime=0.25\GeV^{-2}$ and $s_0=1\GeV^2$, the energy shift $s\Rightarrow \tau s$ results in quite small decrease of the slope, by only $0.35\GeV^{-2}$, which we can neglect. Then the damping factor turns out to be $D_{abs}^{NN}=0.25$.

After integration of the pion fluxes over $q_{1.2}$, the relation (\ref{720}) between the $pp$ and $\pi\pi$ cross sections simplifies,
\beqn
\!\!\!\!\frac{d\sigma(pp\to nXn)}{dz_1dz_2}&=&
F_{\pi^+/p}(z_1)\,
\sigma^{\pi^+\pi^+}_{tot}(\tau s)
\nonumber\\&\times& 
F_{\pi^+/p}(z_2)\,
D_{abs}^{NN}(s,z_1,z_2),
\label{780}
\eeqn
where the absorption corrected and $q$-integrated effective fluxes of pions are presented in Eqs.~(\ref{800})-(\ref{840}).

To maximize the statistics one can include all registered pairs of neutrons into the analysis,
as was done within the Born approximation in Eq.~(\ref{210}),
\beqn
&&\frac{d\sigma(pp\to nXn)_{z>z_{min}}}
{\sigma^{\pi^+\pi^+}_{tot}(\tau s)}\equiv \Phi(\tau)=
\label{850}
\\&=& \!\!\!
\int\limits_{z_{min}}^{z_{max}}\!\!\! 
\frac{dz_1}{1-z_1}
F_{\pi^+/p}(z_1)
F_{\pi^+/p}(z_2)
D_{abs}^{NN}(s,z_1,z_2).
\nonumber\eeqn
The results of the integration, $\Phi(\tau)$, are plotted in Fig.~\ref{fig:PHI} vs $\tau$
by the upper solid and next to it dashed curves, which include and exclude the feed-down corrections respectively.

The calculations have been done so far with the bottom integration limit in (\ref{850})
$z_{min}=0.5$. The corresponding longitudinal momentum transfer is rather large,
creating problems with extrapolation far away from the pion pole, as we already mentioned above. Therefore if experimental statistics allows, one should try to do measurements with possibly larger $z_{min}$. The corresponding total integrated fluxes $\Phi(z)$ calculated with
$z_{min}=0.5-0.9$ are depicted vs $\tau$ in Fig.~\ref{fig:PHI} by solid curves.

\section{Other iso-vector exchanges}\label{reggeons}

Besides the pion, other iso-vector Reggeons contribute to the meson flux in the proton.
The natural spin-parity Reggeons, $\rho$ and $a_2$ have intercepts, $\alpha_R(0)=1/2$, higher than the pion, so should dominate at sufficiently small $1-z$ \cite{kpp}. However, they
are predominantly spin-flip, so vanish in the forward direction. 
The contribution of these Reggeons to the meson spin-flip flux is evaluate and compared with 
the pion term in \ref{rho}.

The unnatural parity $a_1$ reggeon does not flip the nucleon helicity \cite{kane}. It has a low intercept,
so may be important only at small $z$. It was argued in \cite{kpss-AN} that the $a_1$ pole is rather weak, while the dispersion relation for the axial current is dominated by the $\rho-\pi$
cut, which closely imitates the $a_1$ pole. Such an effective $\tilde a_1$ pole is discussed and evaluated in Appendix~\ref{a1}.

The contributions of other iso-vector Reggeons to the meson flux is found relatively small. 
It should be either added to the pion contribution before comparing with data, or
vice versa, extracted from data. In any case, this correction is found to be smaller than the pion contribution.

\section{Summary}

Detecting leading neutrons produced in $pp$ collisions, with the ZDCs installed in the ALICE, ATLAS and CMS experiments at the LHC, provides a unique opportunity to study pion collisions at very high energies, due to the presence of intensive pion fluxes in the colliding protons.
The result, however, is subject to strong absorptive corrections , which is the main objective of the present paper. Thus, experimental data on double neutron production, $pp\to n\,X\,n$ can be treated as a way to study either the pion-pion cross section, provided that the absorption effects are under control, or, vice versa, as a study of the absorptive effects, making a plausible assumption about the pion cross section (like was done in \cite{kpss,kpps-dis}).

Absorptive corrections emerge due to initial/final state interactions of the participating nucleons with the produced hadronic state $X$ (see Fig.~\ref{fig:pi-pi}), as well as among themselves.
Such a classification allows to avoid double-counting. We describe the interaction of the state $X$ in the dipole representation, replacing the multi-hadron state $X$ by a 4-quark color octet-octet dipole, $\{\bar qq\}_8-\{\bar qq\}_8$. If only such an effect of absorption were presented, 
the factorized form, Eq.~(\ref{160}) or (\ref{195}),
would be valid, like in the Born approximation, but with about twice weaker pion fluxes.
In this situation, one could probably talk about renormalized fluxes, like was proposed in \cite{dino}.
However, the interaction of the nucleons with each other breaks down
factorization introducing an overall suppression factor $D(s,z_1,z_2)\approx0.25$ in the absorption corrected cross section, Eq.~(\ref{720}) or (\ref{780}).

A background to the pion exchange comes from other iso-vector Reggeons.
The natural parity, exchange degenerate $\rho$ and $a_2$ contribute to the meson flux, which flips the nucleon helicity, while the unnatural parity axial-vector $a_1$ and its interference with the pion exchange, are added to the helicity conserving meson flux.
Nevertheless, making the plausible assumption that the cross sections of interaction of these mesons with pions and with each other are similar to the pion-pion one,  we found these corrections relatively small.

The calculations presented here unavoidably involve different assumptions and approximations
leading to a theoretical uncertainty of the results. Although the magnitude of such an uncertainty is difficult to evaluate (as usual), basing on the previous experience and available data for neutron production in  hadronic collisions and DIS, we would estimate the accuracy to range within 10-30\%, depending on kinematics. We expect a better accuracy at higher $z_{1,2}$ and smaller $q_{1,2}$, where one approaches the pion pole and eliminates the background from other Reggeons. 

Notice that the $q$-dependent cross section can be currently measured only with the ZDC installed in the ATLAS experiment, while the CDF and ALICE experiment can measure only the cross section integrated over a certain range of $q$.
Our exposed numerical results were calculated as examples for ad hoc kinematics and experimental constraints. We can perform calculations for a concrete experimental set-up upon request.

%%%%%%%%%%%%%%%%%%%%%%%%
 \def\appendix{\par
 \setcounter{section}{0}
\setcounter{subsection}{0}
 \def\thesection{Appendix \Alph{section}}
\def\thesubsection{\Alph{section}.\arabic{subsection}}
\def\theequation{\Alph{section}.\arabic{equation}}
\setcounter{equation}{0}}

 \appendix

  \section{The status of the triple-Regge Phenomenology}\label{3R}
 
 The triple-Regge fit to data large rapidity gap process $pp\to pX$ 
was performed in \cite{kklp}. The cross section of the process $a+b\to c+X$ pictorially presented in Fig.~\ref{fig:3r},
%%%%%%%%%%%%%%%%%%%%%
  \begin{figure}[htb]
\centerline{
  \scalebox{0.4}{\includegraphics{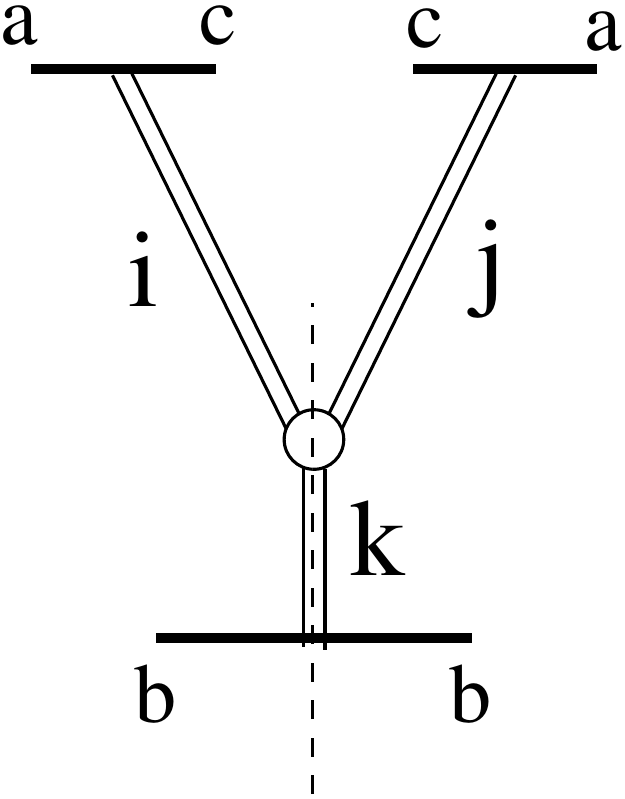}}}
 \caption{The triple-Regge graph presenting the cross section of the process $a+b\to c+X$.
 The Reggeons $i$, $j$ and $k$ can be either the Pomeron $\mathbb P$, or the leading Reggeon $\mathbb R$. The vertical dashed line shows the unitarity cut.
 }
 \label{fig:3r}
 \end{figure}
 %%%%%%%%%%%%%%%%%%%%%
was parametrised as,
 \beqn
 \frac{d\sigma(pp\to pX)}{dx\,dt}\!\!&=&\!\!\!\!
 \sum\limits_{i,j,k=\mathbb{P,R}}
 G_{ijk}(t)\,(1-x)^{\alpha_k(0)-\alpha_i(t)-\alpha_j(t)}
 \nonumber\\ &\times&
 \left(\frac{s}{s_0}\right)^{\alpha_k(0)-1}\!\!\!
 + \left(\frac{d\sigma}{dx\,dt}\right)_{\!\pi\pi\mathbb{P}},
 \label{a100}
 \eeqn
 where diagonal terms have the form,
 \beq
 G_{iik}(t)=G_{iik}(0)\,e^{R_{iik}^2t}.
 \label{a200}
 \eeq
 The off-diagonal $\mathbb{R}$-$\mathbb{P}$ terms were written as,
 \beqn
 &&
 G_{\mathbb{PR}k}(t)+G_{\mathbb{RP}k}(t)=
 2\re G_{\mathbb{PR}k}(t)
  \label{a300} \\&=&
 2\sqrt{2}\,\re G_{\mathbb{PR}k}(0)e^{R_{\mathbb{PR}k}^2t}
\cos\left({\pi\over2}\left[\alpha_{\mathbb{P}}(t)- 
 \alpha_{\mathbb{R}}(t)\right]\right).
\nonumber
 \eeqn
 The last $\pi\pi\mathbb{P}$ term in (\ref{a100}) is calculated with the pion flux Eq.~(\ref{180}), reduced by the isotopic factor $1/2$.
 
 The six parameters fitted in \cite{kklp} to data (first solution) are listed in Table~\ref{solution},
 \begin{table}[htb]
\Caption{
 \label{solution}
The parameters in Eq.~(\ref{a100}) fitted to data on $pp\to pX$ \cite{kklp}.} 
\vspace{-5mm}
  \begin{center}
\begin{tabular}{|c|c|c|c|c|c|c|}
 \hline
 \vphantom{\bigg\vert}  
 &  $G_{\mathbb{PPP}}$
  & $G_{\mathbb{RRP}}$
  & $2\re G_{\mathbb{PRP}}    $              
  & $G_{\mathbb{PPR}} $
  & $G_{\mathbb{RRR}}$
  & $2\re G_{\mathbb{PRR}}$
  \\
\hline &&&&&&\\[-6mm]
   $G_{ijk}(0)$
  &$3.24$
  & $7.2$
  & $6.9 $              
  & $3.2$
  & $5.19$
  & $-9.3$
  \\[-1mm]
   $\bigl(\frac{\mb}{GeV^2}\bigr)$
  &$\pm0.35$
  &$\pm1.9$
  & $\pm1.1 $              
  & $\pm0.6$
  & $\pm7.8$
  & $\pm2.2$
 \\[2mm]
\hline &&&&&&\\[-6mm]
   $R^2_{ijk}$
  &$4.25$
  & $-1.2$
  & $8.5 $              
  & $1.7$
  & $0$
  & $0$
\\[-1mm]
   $\left(GeV^{-2}\right)$
  &$\pm0.24$
  &$\pm0.5$
  & $\pm3.7 $              
  & $\pm0.4$
  & 
  & 
 \\[2mm]
 \hline
  \end{tabular}
\end{center}
\end{table}
%%%%%%%%%%%%%%%%%
 We keep the same values of parameters of the Regge trajectories, as in \cite{kklp}, in particular the Pomeron intercept  $\alpha_{\mathbb{P}}(0)=1$.
 %$\alpha_{\mathbb{P}}^\prime=0.3\GeV^{-2}$; $\alpha_{\mathbb{R}}(0)=0.5$;
 %$\alpha_{\mathbb{R}}^\prime=0.75\GeV^{-2}$. 
These parameters are effective ones, because they include the effects of various types of Regge cuts, in particular absorption corrections, which steeply increase with energy, significantly compensating the rise of the cross section generated by  the  high Pomeron intercept. Of course the choice of
$\alpha_{\mathbb{P}}(0)=1$ was made in \cite{kklp} without a strong justification,
simply because it was the common believe in early days of the Regge theory, but it reproduces data at LHC quite correctly as is demonstrated below.
 
 The TOTEM experiment \cite{totem-sd} found the cross section integrated within the interval of invariant masses $3.4<M_X<1100\GeV$, $\sigma_{sd}=3.25\pm 0.65\mb$, which also agrees with the expected value of $\sigma_{sd}=4.2\mb$. Moreover, the $t$-slope of the differential single-diffractive cross section was measured within different mass intervals, and the results are presented in Table~\ref{slopes}. 
 %%%%%%%%%%
   \begin{table}[htb]
\Caption{
 \label{slopes}
The slope of the single-diffractive cross section averaged over three intervals of invariant masses measured at $\sqrt{s}=7\TeV$ in the TOTEM experiment \cite{totem-sd}, vs calculated with Eq.~(\ref{a100}).}
\vspace{-7mm}
  \begin{center}
\begin{tabular}{|c|c|c|c|}
 \hline
 \vphantom{\bigg\vert}  
 $M_X (GeV)$
 &  $3.4-8$
  & $8-350$
  & $350-1100$              
   \\[-1mm]
\hline &&&\\[-6mm]
   $B\, (\GeV^{-2})$
  &$10.1$
  & $8.5$
  & $6.8 $              
    \\[-2mm]
   {\small TOTEM}
  &&& 
  \\[-1mm]
\hline &&&\\[-6mm]
   $\left\la B\right\ra_{M_X}(\GeV^{-2})$
  &$10.96$
  & $9.06$
  & $7.25 $              
 \\[-2mm]
 {\small  Eq.~(\ref{a100})}
  &  &  &    \\
 \hline
  \end{tabular}
\end{center}
\end{table}
%%%%%%%%%
The measured slopes are compared with calculated and averaged over the corresponding mass intervals,
 \beq
 \left\la B\right\ra_{M_X} =
 \frac{\int_{z_1}^{z_2} dz\,d\sigma_{sd}/dzdt\bigr|_{t=0}}
 {\int_{z_1}^{z_2} dz\,d\sigma_{sd}/dz},
 \label{a500}
 \eeq
 where $z_{1,2}$ correspond to the minimal and maximal value of the invariant mass,
 respectively, $z=1-M_X^2/s$.  The measured values and predictions agree well.

 The CMS experiment \cite{cms-sd} measured the one-side single-diffraction cross section  at $\sqrt{s}=7\TeV$ integrated within $10^{-5.5} <1-z < 10^{-2.5}$ and found, 
 $
 \sigma_{sd}=2.14 \pm 0.02
 \begin{array}{c}+0.33\\
[-2mm] -0.29\end{array}$.
 This agrees well with $\sigma_{sd}=2.43\mb$ predicted by Eq.~(\ref{a100}) with the parameters in Table~\ref{solution}.
 
 The ATLAS experiment \cite{atlas-sd} measured the single-diffractive cross section at $\sqrt{s}=7\TeV$ integrated within $10^{-5.1}<1-z<10^{-3.8}$ at $\sigma_{sd}=1.52\pm0.12\mb$, which is to be compared with  $\sigma_{sd}=1.08\mb$ obtained with Eq.~(\ref{a100}).
 
 The ALICE experiment \cite{alice-sd} measured the single-diffraction cross section at $M_X<200\GeV$ and different energies, as is presented in Table~\ref{alice}.
   \begin{table}[htb]
\Caption{
 \label{alice}
The single-diffractive cross section measured in \cite{alice-sd} for invariant masses $M_X<200\GeV$ at
$\sqrt{s}=0.9,\ 2.76$ and $7\TeV$, in comparison with expectations based on Eq.~(\ref{a100}).}
\vspace{-7mm}
  \begin{center}
\begin{tabular}{|c|c|c|c|}
 \hline
 \vphantom{\bigg\vert}  
 $\sqrt{s}(TeV)$
 &  $0.9$
  & $2.76$
  & $7$              
   \\[-1mm]
\hline &&&\\[-6mm]
   $ \begin{array}{c}
 \sigma^{exp}_{sd} (\mb)\\[-2mm]
 M_X<200\GeV
 \end{array}$
  &$5.6\begin{array}{c}
  +0.8\\[-2mm]
  -1.05
  \end{array}$
   &$6.1\begin{array}{c}
  +1.95\\[-2mm]
  -2.65
  \end{array}$
  &$7.45\begin{array}{c}
  +1.7\\[-2mm]
  -2.95
  \end{array}$
    \\
 \hline &&&\\[-6mm]
   $ \begin{array}{c}
 \sigma^{theor}_{sd} (\mb)\\[-1mm]
Eq.~(\ref{a100})
 \end{array}$
  &$5.13$
  & $4.3$
  & $3.86 $              
\\
 \hline
  \end{tabular}
\end{center}
\end{table}
%%%%%%%%%%%%%
   \begin{table}[htb]
\Caption{
 \label{alice}
The single-diffractive cross section measured at $\sqrt{s}=7\TeV$ in different invariant mass
intervals by the  CMS \cite{cms-sd}, TOTEM \cite{totem-sd} and ATLAS \cite{atlas-sd} experiments. The data are compared with the the results of Eq.~(\ref{a100}).}
\vspace{-7mm}
  \begin{center}
\begin{tabular}{|c|c|c|c|}
 \hline
 \vphantom{\bigg\vert}  
 $M_X[GeV]$
 &  $ \begin{array}{c}
 12.5 - 393.6\\[-2mm]
 ({\rm CMS}) \end{array}$
  &  $ \begin{array}{c}
 3.4 - 1100\\[-2mm]
 ({\rm TOTEM}) \end{array}$ 
  &  $ \begin{array}{c}
9.7 - 88.1\\[-2mm]
 ({\rm ATLAS}) \end{array}$            
   \\[-1mm]
\hline &&&\\[-6mm]
   $ \sigma^{exp}_{sd} [\mb] $
    &$2.14\pm0.02\!\!
    \begin{array}{c}
  +0.33\\[-3mm]
  -0.29
  \end{array}$
   &$3.25\pm0.65$
   &$1.52\pm0.12$
    \\
 \hline &&&\\[-6mm]
   $ \begin{array}{c}
 \sigma^{theor}_{sd} [\mb]\\[-1mm]
Eq.~(\ref{a100})
 \end{array}$
  &$2.43$
  & $4.2$
  & $1.08 $              
\\
 \hline
  \end{tabular}
\end{center}
\end{table}

%%%%%%%%%%%%%
    \begin{table}[htb]
\Caption{
 \label{alice}
The single-diffractive cross section measured in the CDF experiment at the Tevatron \cite{cdf-sd,dino99} 
at $\sqrt{s}=546$ and $1800\GeV$ for $x>0.95$ in comparison with expectations based on Eq.~(\ref{a100}).}
\vspace{-7mm}
  \begin{center}
\begin{tabular}{|c|c|c|}
 \hline
 \vphantom{\bigg\vert}  
 $\sqrt{s}[GeV]$
 &  $546$
  & $1800$
   \\[-1mm]
\hline &&\\[-6mm]
   $ \begin{array}{c}
 \sigma^{exp}_{sd} [\mb]\\[-2mm]
z>0.95
 \end{array}$
  &$4.17\pm 0.18$
   &$4.56\pm 0.23$
    \\
 \hline &&\\[-6mm]
   $ \begin{array}{c}
 \sigma^{theor}_{sd} [\mb]\\[-1mm]
Eq.~(\ref{a100})
 \end{array}$
  &$4.6$
  & $5.24$\\
  \hline 
  \end{tabular}
\end{center}
\end{table}
 
We conclude that the old triple-Regge fit \cite{kklp} successfully predicts the magnitude and $q$-dependence of the single diffraction cross section measured at LHC,
so can be employed for calculation of the feed-down corrections presented in Sect.~\ref{feed-down}.

%\section{Other iso-vector mesons}\label{reggeons}
%\label{classified} \setcounter{equation}{0}

%\section{Appendix}

%\def\appendix{\par
 %\setcounter{section}{0}
 %\setcounter{subsection}{0}
%% \def\thesection{Appendix \Alph{section}}
 %\def\thesubsection{\Alph{subsection}}
 %\def\theequation{\Alph{subsection}.\arabic{equation}}
 %\setcounter{equation}{0}}                  

%\appendix

\section{\boldmath $\rho$ and $a_2$ Reggeons}\label{rho}
%\label{appendA}
\setcounter{equation}{0}

The iso-vector $\rho$ and $a_2$ Reggeons are mostly spin-flip \cite{kane,irving}, so we neglect their small non-flip part in what follows. These Regeon exchanges can be treated as a spin-flip meson flux in the proton, in addition to the pion one, Eq.~(\ref{560}).
\beqn
f_{\rho^+/p}^{(s)}(z,q)=
f_{a_2^+/p}^{(s)}(z,q)=
q^2\,\frac{1-z}{z^2}
\left|\psi^{(s)}_\rho(z,q)\right|^2\!\!.
\label{860}
\eeqn
We rely here on Regge duality, which leads to the exchange degeneracy of $\rho$ and $a_2$,
i.e. equality of their Regge trajectories and $RNN$ vertices.  

Eq.~(\ref{860}) $\psi^\rho_s(q,z)$, compared with Eq.~(\ref{560}), contains the imaginary part neglected for pions,
and several other modifications \cite{kpps-dis}, 
\beqn
\psi^\rho_s(z,q)=
\frac{\Omega_\rho(z)}{2\pi \,q\, \beta_\rho^3}
\int\limits_0^\infty db\,b\,J_1(bq)
K_1(b/\beta_\rho)\,S^{\pi N}_{abs}(b,z).
\label{880}
\eeqn
 Here we made a natural assumption that $S^{\rho N}=S^{\pi N}_{abs}$, and
  \beqn
\Omega_\rho(z) &=&\frac{\pi\,\alpha_\rho^\prime}{4}g_{\rho^+pn}\eta_\rho(0)
z(1-z)^{-\alpha_\rho(0)+\alpha^\prime_\rho q_L^2}
e^{-R_\rho^2 q_L^2};
\nonumber\\
\beta_\rho^2&=&{1\over z}\,\left[
R_\rho^2-\alpha_\rho^\prime\,\ln(1-z)\right];
\label{900}
 \eeqn
with $\eta_\rho(0)=-i-1$. 
 $\Omega_\rho(z)$ contains an additional $z$-dependence, a factor $\sim1/\sqrt{1-z}$, compared to 
the pion exchange, Eq.~(\ref{560}), because the $\rho$ intercept is higher.

 For the vertex function $G_{\rho NN}(t)=g_{\rho NN}\exp(R_{\rho}^2t)$ we rely on the phenomenological global Regge analysis \cite{irving} of high-energy hadronic data, which resulted in $g_{\rho NN}= 0.5\, g_{\pi NN}$, and $R^2_\rho=1\GeV^{-2}$.

Notice that the fluxes of pions, $\rho$ and $a_2$, can be added without interferences, which are suppressed as $1/(\tau s)$ because the quantum numbers of these mesons do not allow diffractive $\pi\to\rho$ transitions.

Thus, the $\rho$ and $a_2$ can be added to the spin-flip flux of the pions, Eq.~(\ref{560}), and then the overall absorption factor $S_{abs}^{NN}$ should be applied, as described above.

Correspondingly, $\rho$ contributes to the $q$-integrated spin-flip meson flux as,
 \beqn
F^{(s)}_{\rho^+/p}(z)&=&
\frac{(1-z)
\Omega^2_\rho(z)}{4\pi^3z^2\,\beta_\rho^6}
\int d^2b\,K_1^2\!\left({b\over\beta_\rho}\right)
\nonumber\\ &\times&
\left[S^{\pi N}_{abs}(b,z)\right]^2,
\label{910}
 \eeqn
and the same amount comes from $a_2$.

Fig.~\ref{fig:pi-rho} presents the result of calculations for the $q$-integrated spin-flip flux, the pion term, in comparison with the $\rho$ and $a_2$ contributions, and the full flux.
%%%%%%%%%%%%%%%%%%%%%
  \begin{figure}[htb]
\centerline{
  \scalebox{0.3}{\includegraphics{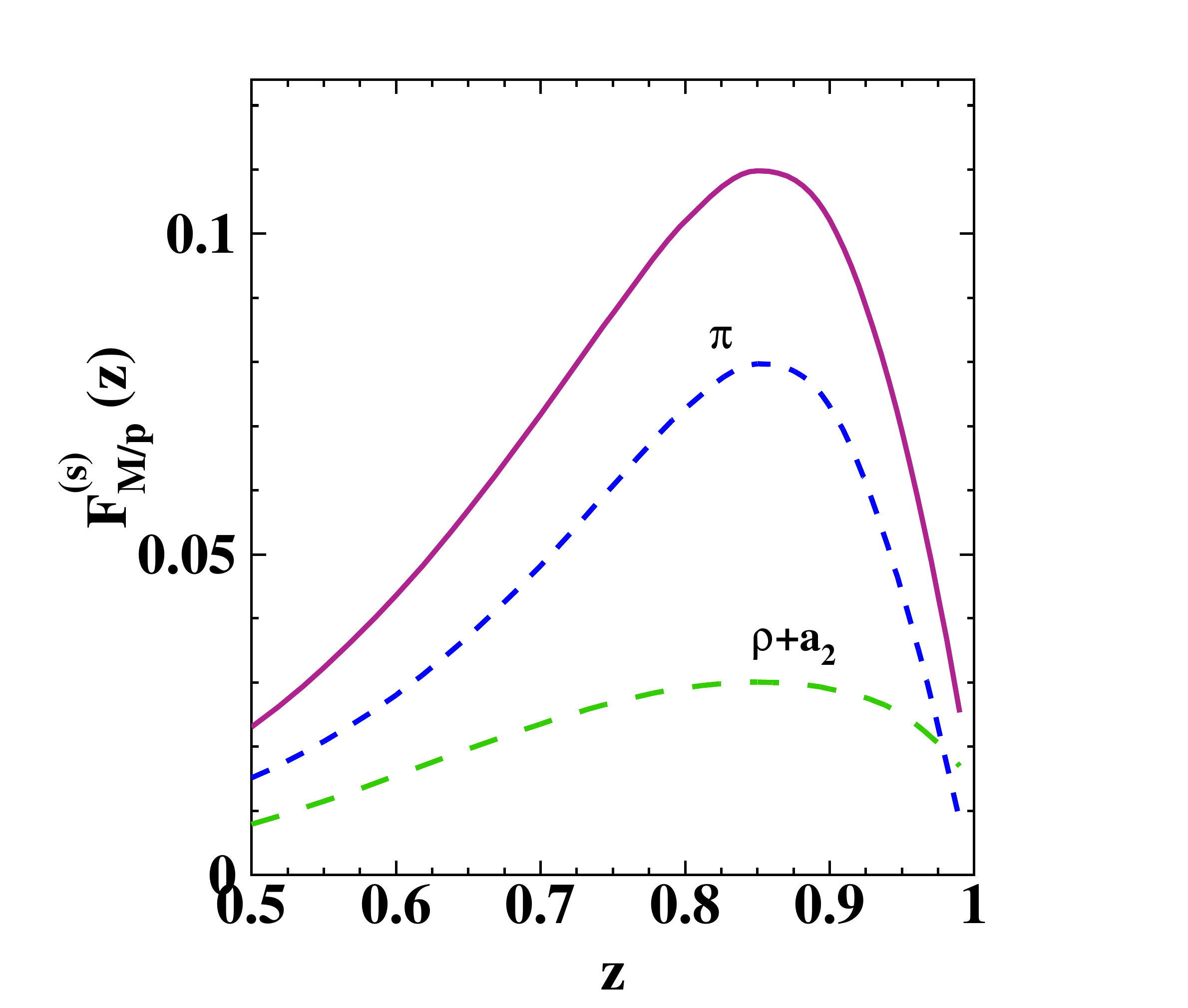}}}
 \caption{(Color online) The $q$-integrated spin-flip meson flux $F_{M/p}^{(s)}(z)$, including pion (upper dashed) and $\rho$ and $a_2$ mesons (bottom dashed) contributions. The solid curve shows the full flux.
}
 \label{fig:pi-rho}
 \end{figure}
 %%%%%%%%%%%%%%%%%%%%%
The pion term dominates within the presented interval of $z$, except at small and very large $z$. Notice that these Reggeons are relatively enhanced at large $q$ because have much smaller $t$-slope than pion. Besides, they are spin-flip, so do not contribute at $q=0$
and are much suppressed in the meson flux integrated up to certain values of $q_{max}$ as was done above in Sect.~\ref{S1,2}.

%%%%%%%%%%%%%%%%%%%%%%%%

\section{\boldmath Effective $\tilde a_1$ Reggeon}\label{a1}
\setcounter{equation}{0}

The unnatural spin-parity $a_1$ exchange is predominantly spin non-fip \cite{kane}, so it should contribute the meson flux $f_{M/p}^{(0)}$. Moreover, it interferes with the pion flux, because the diffractive cross section $\pi p\to a_1 p$ is nearly energy independent.

However, in the dispersion relation for the axial current the $a_1$ pole is a weak singularity. Differently from the vector current, dominated by the $\rho$-pole, analogous assumption for the axial current leads to a dramatic contradiction with data \cite{pik-stod,marage}. 
It was proposed in \cite{belkov,marage} to replace the $a_1$ pole by an effective one $\tilde a_1$, which gets the main contribution from the $\rho$-$\pi$ cut,
located in the complex $Q^2$ plane close to the $a_1$ pole, and which can imitate it. 
Indeed, data on diffractive production $\pi\to\rho\pi$ in in the $1^+S$ wave show a strong and narrow peak near the $a_1$ mass \cite{kpss-AN}, formed by the Deck effect.
Inclusion of such an effective Reggeon
$\tilde a_1$, representing the $\rho$-$\pi$ cut, into the calculation of the single-spin asymmetry of leading neutrons \cite{kpss-AN}, and neutron  
production in deep-inelastic scattering \cite{kpps-dis}, led to a good agreement with data without adjustment of any parameters.

The main features of the effective $\tilde a_1$ Reggeon are as follows \cite{kpss-AN}.
The Regge trajectory, corresponding to the $\rho$-$\pi$ cut, is,
\beq
\alpha_{\tilde a_1}(t)=\alpha_{\pi\rho}(t)=\alpha_\pi(0)+\alpha_\rho(0)-1+
\frac{\alpha_\pi^\prime\alpha_\rho^\prime}
{\alpha_\pi^\prime+\alpha_\rho^\prime}\,t ,
 \label{920} 
 \eeq 
so $\alpha_{\tilde a_1}(0)=-0.5$; $\alpha_{\tilde a_1}^\prime=0.45\GeV^{-2}$.
Correspondingly, the phase factor reads,
\beq
\eta_{\tilde a_1}(t)=-i-\tan\!\left[\frac{\pi\alpha_{\tilde a_1}(t)}{2}\right].
\label{940}
\eeq
Notice that the Regge trajectory Eq.~(\ref{920}) crosses $\alpha_{\tilde a_1}(t_0)=-1$
at $t=t_0=-1/(2\alpha_{\tilde a_1}^\prime)$, which is within the kinematic interval under consideration. In this point the signature factor Eq.~(\ref{940}) has a ghost  pole, which must be compensated by a zero in the residue function, in order to eliminate the wrong signature nonsense pole from the Regge trajectory \cite{irving,collins}. So we introduce into the amplitude an additional factor,
$
\gamma(t)=1+t/t_0
$,
which eliminates the wrong pole and doesn't affect much the amplitude far away from it.
Expanding the real part in (\ref{940}) near the pole we get,
\beq
\gamma(t)\, \eta_{\tilde a_1}(t)\approx
\frac{2}{\pi\alpha_{\tilde a_1}^\prime t_0}.
\label{950}
\eeq
Here we are focused on the small-$z$ region, where the $\tilde a_1$ contribution is much enhanced due to its low Regge intercept. In this region one approaches the ghost pole in the real part and can neglect the relatively small imaginary part. This is not be an accurate approximation at large $z\to1$ and small $q$, where the $\tilde a_1$ contribution is very small anyway (see Fig.~\ref{fig:a1-pi}).
In this region of $z$ the real part has no singularity, and the imaginary part might be essential, like in single-spin asymmetry of neutrons, measured in \cite{phenix-neutrons1,phenix-neutrons2} at $z>0.8$. In this case, one should rely on the phase factor Eq.~(\ref{940}), as was done in \cite{kpss-AN}, rather than on the approximation
(\ref{950}).

The $\tilde a_1NN$ vertex is parametrized as $G_{\tilde a_1pn}(t)=g_{\tilde a_1pn}\exp(R_{\tilde a_1}^2t)$. We fix the radius at $R_{\tilde a_1}^2=R_\rho^2=1\GeV^{-2}$,
because $\rho$ and $a_1$ are the chiral partners.
The ${\tilde a_1}NN$ coupling was evaluated in \cite{kpss-AN} based on PCAC and the 
second Weinberg sum rule, in which the spectral functions of the vector and axial
currents are represented by the $\rho$ and the effective ${\tilde a_1}$ poles respectively.
This allows to fix the ${\tilde a_1}NN$ coupling at,
$
g_{\tilde a_1 NN}/g_{\pi NN}\approx 0.5.
$
The $\tilde a_1$ contributed to the spin non-flip flux of mesons, in addition to the pion one, Eq.~(\ref{540}),
\beq
f_{\tilde a_1/p}^{(0)}(z,q) =
\frac{(1-z)}{z}\,q_L^2\,
\left|\psi^{(0)}_{\tilde a_1}(z,q)\right|^2.
\label{960}
\eeq

In the Born approximation  $\psi^{(0)}_{\tilde a_1}(z,q)$ reads \cite{kpss-AN}
 \beqn
\psi^{B(0)}_{\tilde a_1}(z,q) &=&
{1\over8}\,\xi(z,q)\alpha_{\tilde a_1}^\prime
(\gamma\,\eta_{\tilde a_1})
\nonumber\\ &\times&
G_{\tilde a_1 pn}(t)
(1-z)^{-\alpha_{\tilde a_1}(t)},
\label{970}
 \eeqn
while $\psi^{(0)}_{\tilde a_1}$, corrected for absorptive factors $S^{\tilde a_1 N}(b_{1,2},z_{1,2})$ (assumed to be the same as $S^{\pi N}_{abs}$),
has the form,
\beqn
\psi^{(0)}_{\tilde a_1}(z,q) &=& \xi(z,q)\,
\frac{z\,\Omega_{\tilde a_1}(z)}{2\pi R_{\tilde a_1}^2}
\int\limits_0^\infty db\,b\,J_0(bq)
\nonumber\\ &\times&
K_0(\sqrt{z}\,b/R_{\tilde a_1})\,S^{\pi N}_{abs}(b,z).
\label{980}
\eeqn
The coefficient 
\beq
\xi(z,q)=\frac{2m_N}{\sqrt{|t|}},
\label{990}
\eeq
is related to the spin structure of the axial-vector-nucleon vertex, $e^L_\mu\bar n\gamma_5\gamma_\mu p$ compared with the pion preudo-scalar vertex $\bar n\gamma_5 p$ \cite{kpss-AN,kpps-dis}; and
  \beqn
\Omega_{\tilde a_1}(z) &=&\frac{\pi\alpha_{\tilde a_1}^\prime}{4}\,g_{{\tilde a_1}pn}\,(\gamma\,\eta_{\tilde a_1})\,
e^{-R_{\tilde a_1}^2 q_L^2}
\nonumber\\ &\times&
(1-z)^{-\alpha_{\tilde a_1}(0)+\alpha^\prime_{\tilde a_1}q_L^2}.
\label{1000}
 \eeqn

Now we should correct the term $f_{\pi^+/p}^{(0)}(z,q)$ in the meson spin non-flip flux in Eq.~(\ref{720}), adding the $\tilde a_1$ contribution,
\beqn
f_{M/p}^{(0)}(z,q) &=&
\frac{1-z}{z}\,q_L^2
\Biggl[
\left|\psi^{(0)}_\pi(q,z)\right|^2 +
\left|\psi^{(0)}_{\tilde a_1}(q,z)\right|^2 
\nonumber\\ &+&
2\re\, \psi^{(0)}_{\pi {\tilde a_1}}(q,z)\Biggr].
\label{1020}
\eeqn
As far as both the pion and $\tilde a_1$ amplitudes are real, the interference term reads,
 \beq
 2\re\, \psi^{(0)}_{\pi {\tilde a_1}}(q,z)= 
 2\kappa\,
\psi^{(0)}_\pi(q,z)\,
 \psi^{(0)}_{\tilde a_1}(q,z),
  \label{1040}
 \eeq
where the factor $\kappa$ controls the relative magnitude of the interference term, 
\beq
\kappa=\sqrt{
\frac{d\sigma(\pi p\to\pi\rho p)/dp_T^2}
{d\sigma(\pi p\to\pi p)/dp_T^2}\Bigr|_{p_T=0}}
= 0.29,
\label{1060}
\eeq
%%%%%%%%%%%%%%%%%%%%%
  \begin{figure}[htb]
\centerline{
\scalebox{0.3}{\includegraphics{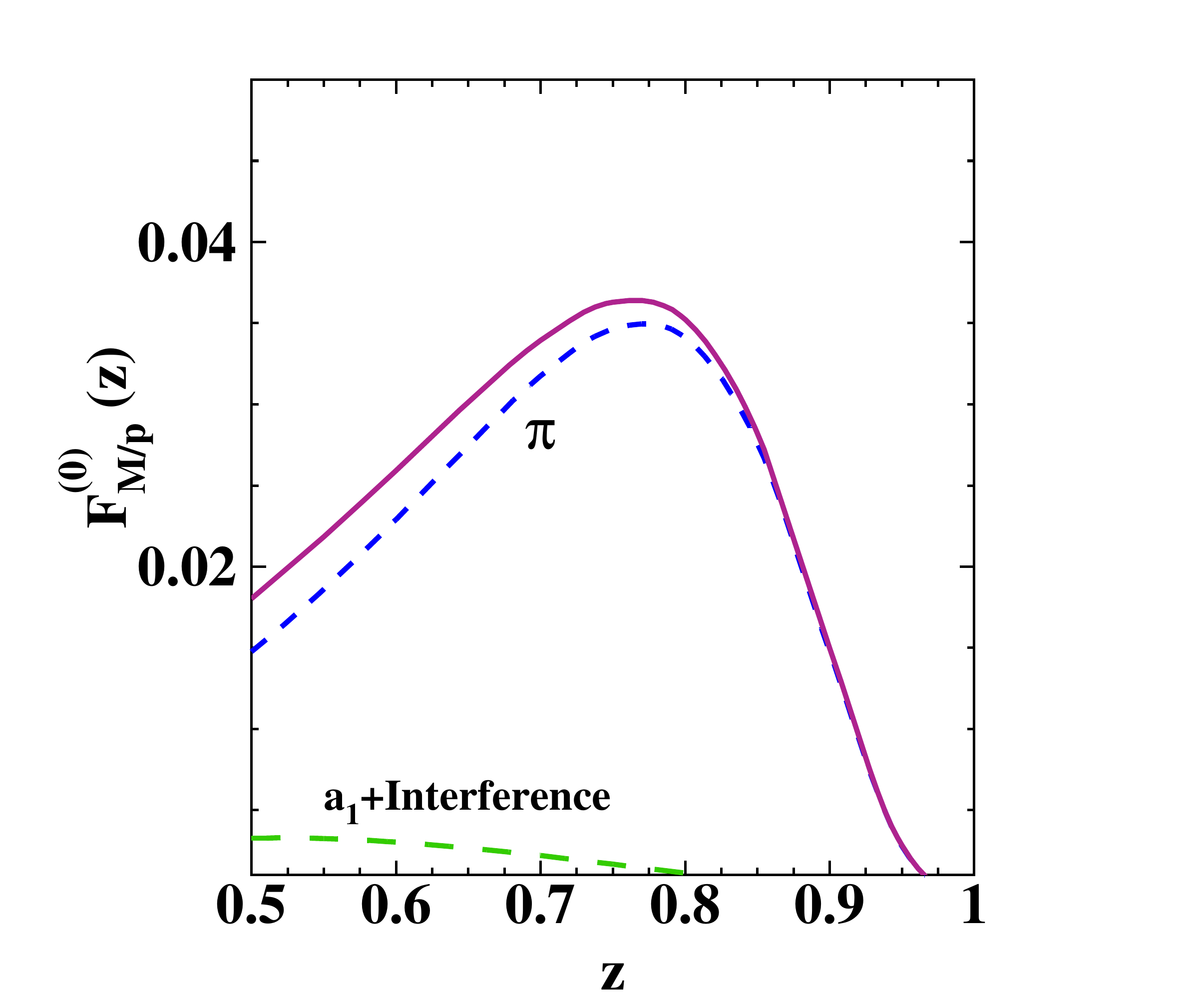}}}
 \caption{(Color online) The $q$-integrated spin non-flip meson flux $F_{M/p}^{(0)}(z)$, including pion (upper dashed) and the combined contribution of $\tilde a_1$ exchange and its interference with pion (bottom dashed). The solid curve shows the full meson flux concerving helicity.
}
 \label{fig:a1-pi}
 \end{figure}
 %%%%%%%%%%%%%%%%%%%%%

This was evaluated in \cite{kpss-AN} from data in $\pi$-$p$ collisions at c.m. energy squared $150\GeV^2$. Although data for single diffraction at high energies agree with energy independence, these are absorptive corrections to the single-diffractive cross section. 
Since we already corrected for absorption to the contributions of $\pi$ and $\tilde a_1$, to avoid double counting we should employ a net diffraction $\pi\to\tilde a_1$, i.e. without absorptive corrections. Then the diffractive cross section is expected to have the same energy dependence as elastic, resulting in an energy independent $\kappa$, Eq.~(\ref{1060}).

The contribution to the $q$-integrated meson flux conserving helicity of the effective Regge pole $\tilde a_1$ and its interference with the pion exchange are plotted in Fig.~\ref{fig:a1-pi}, in comparison with the pion exchange.

\begin{acknowledgments}

This work was supported in part
by Fondecyt (Chile) grants 1130543, 1130549, 1140390,
and by the ECOS grant C12E04.
\end{acknowledgments}


\begin{thebibliography}{99}



 \bibitem{totem} 
  G.~Antchev,  {\it et al.} [TOTEM Collaboration],
  %``First measurement of the total proton-proton cross section at the LHC energy of {\surd} s =7 TeV,''
  Europhys.\ Lett.\  {\bf 96}, 21002 (2011);
   Europhys.\ Lett.\  {\bf 95}, 41001 (2011).
    
\bibitem{alpha} 
  G.~Aad {\it et al.}  [ATLAS Collaboration],
  %``Measurement of the total cross section from elastic scattering in $pp$ collisions at $\sqrt{s}=7$ TeV with the ATLAS detector,''
  arXiv:1408.5778 [hep-ex].

\bibitem{pdg}  K.A. Olive et al. (Particle Data Group), Chin. Phys. C, 38, 090001 (2014). 

\bibitem{review} 
  J.~Gasser,
  %``On the history of pion-pion scattering,''
  PoS EFT {\bf 09}, 029 (2009).

\bibitem{landshoff} 
  A.~Donnachie and P.~V.~Landshoff,
  %``Hard Diffraction: Production of High p(T) Jets, W or Z, and Drell-Yan Pairs,''
  Nucl.\ Phys.\ B {\bf 303}, 634 (1988).

\bibitem{gotsman} 
  E.~Gotsman, E.~Levin and U.~Maor,
  %``A model for strong interactions at high energy based on the CGC/saturation approach,''
  arXiv:1408.3811 [hep-ph].
  
  \bibitem{soffer} 
  C.~Bourrely, J.~Soffer and T.~T.~Wu,
  %``Do we understand near-forward elastic scattering up to TeV energies?,''
  arXiv:1405.6698 [hep-ph].
  
  \bibitem{erasmo} 
  E.~Ferreira, T.~Kodama and A.~K.~Kohara,
  %``Elastic Amplitudes and Observables in pp Scattering,''
  arXiv:1411.3518 [hep-ph].
  
  \bibitem{martin} 
  A.~D.~Martin, M.~G.~Ryskin and V.~A.~Khoze,
  %``From hard to soft high-energy pp interactions,''
  arXiv:1110.1973 [hep-ph].
  
  \bibitem{kaidalov} 
  A.~B.~Kaidalov and M.~G.~Poghosyan,
  %``Description of soft diffraction in the framework of reggeon calculus: Predictions for LHC,''
  arXiv:0909.5156 [hep-ph].

\bibitem{ostapchenko} 
  S.~Ostapchenko,
  %``Total and diffractive cross sections in enhanced Pomeron scheme,''
  Phys.\ Rev.\ D {\bf 81}, 114028 (2010).
  
\bibitem{kst2} 
  B.~Z.~Kopeliovich, A.~Sch\"afer and A.~V.~Tarasov,
  %``Nonperturbative effects in gluon radiation and photoproduction of quark pairs,''
  Phys.\ Rev.\ D {\bf 62}, 054022 (2000).  
  
\bibitem{k3p} 
  B.~Z.~Kopeliovich, I.~K.~Potashnikova, B.~Povh and E.~Predazzi,
  %``Nonperturbative gluon radiation and energy dependence of elastic scattering,''
  Phys.\ Rev.\ Lett.\  {\bf 85}, 507 (2000);
  %``Soft QCD dynamics of elastic scattering in impact parameter representation,''
  Phys.\ Rev.\ D {\bf 63}, 054001 (2001).

\bibitem{kpp2012} 
  B.~Z.~Kopeliovich, I.~K.~Potashnikova and B.~Povh,
  %``Two-scale hadronic structure and elastic pp scattering: predicted and measured,''
  Phys.\ Rev.\ D {\bf 86}, 051502 (2012).

\bibitem{pir-1} 
  A.~I.~Shoshi, F.~D.~Steffen and H.~J.~Pirner,
  %``S matrix unitarity, impact parameter profiles, gluon saturation and high-energy scattering,''
  Nucl.\ Phys.\ A {\bf 709}, 131 (2002).

\bibitem{zeus}
  S.~Chekanov {\it et al.}  [ZEUS Collaboration],
  Nucl.\ Phys.\  B {\bf 637}, 3 (2002);
  Nucl.\ Phys.\  B {\bf 776}, 1 (2007).

\bibitem{h1} 
  F.~D.~Aaron {\it et al.}  [H1 Collaboration],
  %``Measurement of Leading Neutron Production in Deep-Inelastic Scattering at HERA,''
  Eur.\ Phys.\ J.\ C {\bf 68}, 381 (2010).
  
\bibitem{kpps-dis} 
  B.~Z.~Kopeliovich, I.~K.~Potashnikova, B.~Povh and I.~Schmidt,
  %``Pion structure function at small x from DIS data,''
  Phys.\ Rev.\ D {\bf 85}, 114025 (2012).

  \bibitem{ap}
  U.~D'Alesio and H.~J.~Pirner,
  Eur.\ Phys.\ J.\  A {\bf 7}, 109 (2000).

\bibitem{strong2} V.A. Khoze, A.D. Martin and M.G. Ryskin, Eur. Phys. J.
C{\bf 18}, 167 (2000).

\bibitem{kkmr}
 A.~B.~Kaidalov, V.~A.~Khoze, A.~D.~Martin and M.~G.~Ryskin,
  Eur.\ Phys.\ J.\  C {\bf 47}, 385 (2006).

\bibitem{kmr} 
  V.~A.~Khoze, A.~D.~Martin and M.~G.~Ryskin,
  %``Information from leading neutrons at HERA,''
  Eur.\ Phys.\ J.\ C {\bf 48}, 797 (2006).
    
     \bibitem{kpss}
  B.~Z.~Kopeliovich, I.~K.~Potashnikova, I.~Schmidt and J.~Soffer,
  %``Damping of forward neutrons in $p p$ collisions,''
  Phys.\ Rev.\  D {\bf 78}, 014031 (2008).

\bibitem{phenix-neutrons1} 
A.~Adare {\it et al.}  [PHENIX Collaboration],
  %``Inclusive cross section and single transverse spin asymmetry for very forward neutron production in polarized p+p collisions at s=200\UTF{2009}\UTF{2009}GeV,''
  Phys.\ Rev.\ D {\bf 88}, no. 3, 032006 (2013).

\bibitem{phenix-neutrons2} 
  Y.~Goto [PHENIX Collaboration],
  %``Inclusive cross section and single transverse-spin asymmetry of very forward neutron production at PHENIX,''
  Phys.\ Part.\ Nucl.\  {\bf 45}, 79 (2014).

  \bibitem{isr} W. Flauger and F. M\"onnig,
Nucl. Phys. B{\bf 109} (1976) 347.
  
\bibitem{hanlon1}
J.~Hanlon, A.~Brody, R.~Engelmann, T.~Kafka, H.~Wahl, A.~A.~Seidl, W.~S.~Toothacker and J.~C.~Van der Velde {\it et al.},
  %``The Inclusive Reactions p n ---> p X and pi+ n ---> p X at 100-GeV/c,''
  Phys.\ Rev.\ Lett.\  {\bf 37}, 967 (1976).

\bibitem{hanlon2}
J J.~Hanlon, A.~Brody, T.~Kafka, S.~Sommars, J.~E.~A.~Lys, C.~T.~Murphy, S.~J.~Barish and S.~Dado {\it et al.},
  %``Slow Proton Production from Neutron Targets at 100-GeV/c and 400-GeV/c,''
  Phys.\ Rev.\ D {\bf 20}, 2135 (1979).

\bibitem{petrov} 
  V.~A.~Petrov, R.~A.~Ryutin and A.~E.~Sobol,
  %``LHC as pi p and pi pi Collider,''
  Eur.\ Phys.\ J.\ C {\bf 65}, 637 (2010);
  {\sl ibid} {\bf 71}, 1667 (2011).
  
  \bibitem{petrov2} 
  A.~E.~Sobol, R.~A.~Ryutin, V.~A.~Petrov and M.~Murray,
  %``Elastic $\pi^{+}p$ and $\pi^{+}\pi^{+}$ scattering at LHC,''
  Eur.\ Phys.\ J.\ C {\bf 69}, 641 (2010).
  
\bibitem{kpp}
  B.~Kopeliovich, B.~Povh and I.~Potashnikova,
  Z.\ Phys.\  C {\bf 73}, 125 (1996).

\bibitem{kaidalov1} G.G. Arakelyan and K.G. Boreskov, Sov. J. Nucl. Phys. {\bf 30},
840 (1979); {\bf 31}, 819 (1980).

\bibitem{kaidalov2} G.G. Arakelyan, K.G. Boreskov and A.B. Kaidalov, Sov. J.
Nucl. Phys. {\bf 33}, 247 (1981).

 \bibitem{kpss-AN} 
  B.~Z.~Kopeliovich, I.~K.~Potashnikova, I.~Schmidt and J.~Soffer,
  %``Single transverse spin asymmetry of forward neutrons,''
  Phys.\ Rev.\ D {\bf 84}, 114012 (2011).
  
  \bibitem{bdmps} 
  R.~Baier, Y.~L.~Dokshitzer, S.~Peigne and D.~Schiff,
  %``Induced gluon radiation in a QCD medium,''
  Phys.\ Lett.\ B {\bf 345}, 277 (1995).
  
  \bibitem{marage}
  B.~Z.~Kopeliovich and P.~Marage,
  %``Low Q**2, high neutrino nu physics (CVC, PCAC, hadron dominance),''
  Int.\ J.\ Mod.\ Phys.\  A {\bf 8}, 1513 (1993).

\bibitem{kklp} 
  Y.~M.~Kazarinov, B.~Z.~Kopeliovich, L.~I.~Lapidus and I.~K.~Potashnikova,
  %``Triple Regge Phenomenology in the Reaction p + p --> p + x,''
  Sov.\ Phys.\ JETP {\bf 43}, 598 (1976)
  [Zh.\ Eksp.\ Teor.\ Fiz.\  {\bf 70}, 1152 (1976)].

  \bibitem{kane}
  H.~E.~Haber and G.~L.~Kane,
  %``The Search for the a1 Meson,''
  Nucl.\ Phys.\  B {\bf 129}, 429 (1977).

  \bibitem{dino} K.~Goulianos, Phys. Lett. B{\bf 358} (1995) 379.
  
  \bibitem{totem-sd} M.~Csan\'ad,
  %``Diffraction measurements at the LHC,''
  arXiv:1312.3803 [hep-ex].

  \bibitem{cms-sd} CMS Collaboration, CMS-PAS-FSQ-12-005.
  
\bibitem{atlas-sd} 
  G.~Aad {\it et al.}  [ATLAS Collaboration],
  %``Rapidity gap cross sections measured with the ATLAS detector in $pp$ collisions at $\sqrt{s}=7$ TeV,''
  Eur.\ Phys.\ J.\ C {\bf 72}, 1926 (2012).

\bibitem{alice-sd} 
  B.~Abelev {\it et al.}  [ALICE Collaboration],
  %``Measurement of inelastic, single- and double-diffraction cross sections in proton--proton collisions at the LHC with ALICE,''
  Eur.\ Phys.\ J.\ C {\bf 73}, no. 6, 2456 (2013).

  \bibitem{cdf-sd} 
  F.~Abe {\it et al.}  [CDF Collaboration],
  %``Measurement of $\bar{p}p$ single diffraction dissociation at $\sqrt{s} = 546$ GeV and 1800 GeV,''
  Phys.\ Rev.\ D {\bf 50}, 5535 (1994).
  
   \bibitem{dino99} 
 K.~A.~Goulianos and J.~Montanha,
  %``Factorization and scaling in hadronic diffraction,''
  Phys.\ Rev.\ D {\bf 59}, 114017 (1999).
  
  \bibitem{irving} 
  A.~C.~Irving and R.~P.~Worden,
  %``Regge Phenomenology,''
  Phys.\ Rept.\  {\bf 34}, 117 (1977).

\bibitem{pik-stod}
  C.~A.~Piketty and L.~Stodolsky,
  %``Diffraction model of high-energy leptonic interactions,''
  Nucl.\ Phys.\  B {\bf 15}, 571 (1970).

\bibitem{belkov}
  A.~A.~Belkov and B.~Z.~Kopeliovich,
  %``ADLER RELATION AND NEUTRINO PRODUCTION OF SINGLE HADRONS,''
  Sov.\ J.\ Nucl.\ Phys.\  {\bf 46}, 499 (1987) 
  [Yad.\ Fiz.\  {\bf 46}, 874 (1987)].

\bibitem{collins} 
  P.~D.~B.~Collins,
   ``An Introduction to Regge Theory and High-Energy Physics,''
  Cambridge 1977, 445p.


\end{thebibliography}
\end{document}